\documentclass[aps,prl,twocolumn,groupedaddress,amsmath,nofootinbib]{revtex4-1}

\usepackage{amssymb}
\usepackage{amsfonts}
\usepackage{amssymb}
\usepackage{amsmath}
\usepackage[dvips]{graphicx}
\usepackage{color}

\usepackage{multirow}

\usepackage{txfonts}
\usepackage{hyperref}

\begin{document}

\title{Characterizing the multipartite continuous-variable entanglement structure from squeezing coefficients and the Fisher information}

\author{Zhongzhong Qin$^{1,2}$}
\author{Manuel Gessner$^{3}$}
\author{Zhihong Ren$^{1,2,4}$}
\author{Xiaowei Deng$^{1,2}$}
\author{Dongmei Han$^{1,2}$}
\author{Weidong Li$^{1,2,4}$}
\email{wdli@sxu.edu.cn}
\author{Xiaolong Su$^{1,2}$}
\email{suxl@sxu.edu.cn}
\author{Augusto Smerzi$^{3,5}$}
\author{Kunchi Peng$^{1,2}$}

\affiliation{$^1$State Key Laboratory of Quantum Optics and Quantum Optics Devices,
Institute of Opto-Electronics, Shanxi University, Taiyuan 030006, People's
Republic of China\\
$^2$Collaborative Innovation Center of Extreme Optics, Shanxi University,
Taiyuan, Shanxi 030006, People's Republic of China\\
$^3$QSTAR, INO-CNR and LENS, Largo Enrico Fermi 2, I-50125 Firenze, Italy\\
$^4$Institute of Theoretical Physics and Department of Physics, Shanxi University, 030006 Taiyuan, China\\
$^5$Institute of Laser Spectroscopy, Shanxi University, Taiyuan 030006, China}
\date{\today}

\begin{abstract}
Understanding the distribution of quantum entanglement over many parties is a fundamental challenge of quantum physics and is of practical relevance for several applications in the field of quantum information. The Fisher information is widely used in quantum metrology since it is related to the quantum gain in metrology measurements. Here we use methods from quantum metrology to microscopically characterize the entanglement structure of multimode continuous-variable  states in all possible multi-partitions and in all reduced distributions. From experimentally measured covariance matrices of Gaussian states with 2, 3, and 4 photonic modes with controllable losses, we extract the metrological sensitivity as well as an upper separability bound for each partition. An entanglement witness is constructed by comparing the two quantities. Our analysis demonstrates the usefulness of these methods for continuous-variable systems and provides a detailed geometric understanding of the robustness of cluster-state entanglement under photon losses.
\end{abstract}

\maketitle

\section*{Introduction}

Entanglement plays a central role in quantum information science \cite{NielsenChuang,Braunstein,Weedbrook}, in particular for quantum computation \cite{Walther2005,Ukai2011,SuNC} and quantum metrology \cite{GiovannettiReview}. An efficient analysis of the quantum resources for such applications requires a detailed understanding of the correlation structure of multipartite quantum states and the development of experimentally feasible methods for their experimental characterization \cite{LuPRX}.

Entanglement of continuous-variable (CV) systems has been studied intensively over the past years \cite{Braunstein,Weedbrook}. The most common method for the analysis of bi-partitions is the positive partial transposition (PPT) criterion, which is highly efficient and easy to implement for Gaussian states \cite{PPTCriteria,DuanCriteria}. Providing a microscopic picture of the entanglement structure in terms of all possible combinations of subsystems, i.e., multi-partitions, is a considerably more difficult task \cite{GuehneT}. Multipartite CV entanglement criteria for specific partitions can be derived from uncertainty relations \cite{LoockCriteria} or by systematic construction of entanglement witnesses \cite{Sperling}. While criteria of this kind are experimentally convenient in many cases \cite{Su2007,Chen,Furusawa,Gerke2015}, they require the additional effort of determining the separability bound as a function of the observables at hand, which can be a complicated problem in general. Moreover, abstract entanglement witnesses usually provide little intuition about the physical significance and origin of the entanglement.

The Fisher information relates the multipartite entanglement between the subsystems to the sensitivity for quantum parameter estimation \cite{GessnerPRA2016}. This approach has proven to be extremely successful with discrete-variable systems, especially for spin systems of cold atoms \cite{LucaRMP}. The Fisher information can furthermore be efficiently approximated for Gaussian spin states by means of experimentally convenient spin squeezing coefficients \cite{Wineland92,Sorensen}. Using these methods, multipartite entanglement of large numbers of particles has been demonstrated by collective measurements \cite{LucaRMP,Lucke,Strobel,Bohnet}.

An extension of the theoretical framework to CV systems has been achieved recently by combining the quantum Fisher information with local variances \cite{GessnerPRA2016} and the development of a bosonic multi-mode squeezing coefficient \cite{GessnerQuantum}. The squeezing coefficient is based on a second-order approximation of the quantum Fisher information and represents an easily accessible entanglement criterion. A microscopic understanding of the inseparability properties in all possible partitions of the system is provided by the information from local measurements on the subsystems. Local observables are routinely measured in CV systems, such as photonic cluster states \cite{Su2007,Seiji,Furusawa}. The separability bounds for the metrological sensitivity are directly obtained from the local data and need not be determined theoretically. Entanglement criteria based on the quantum Fisher information further provide a geometric interpretation in phase space.

\begin{figure*}[tp]
\includegraphics[width=.95\textwidth]{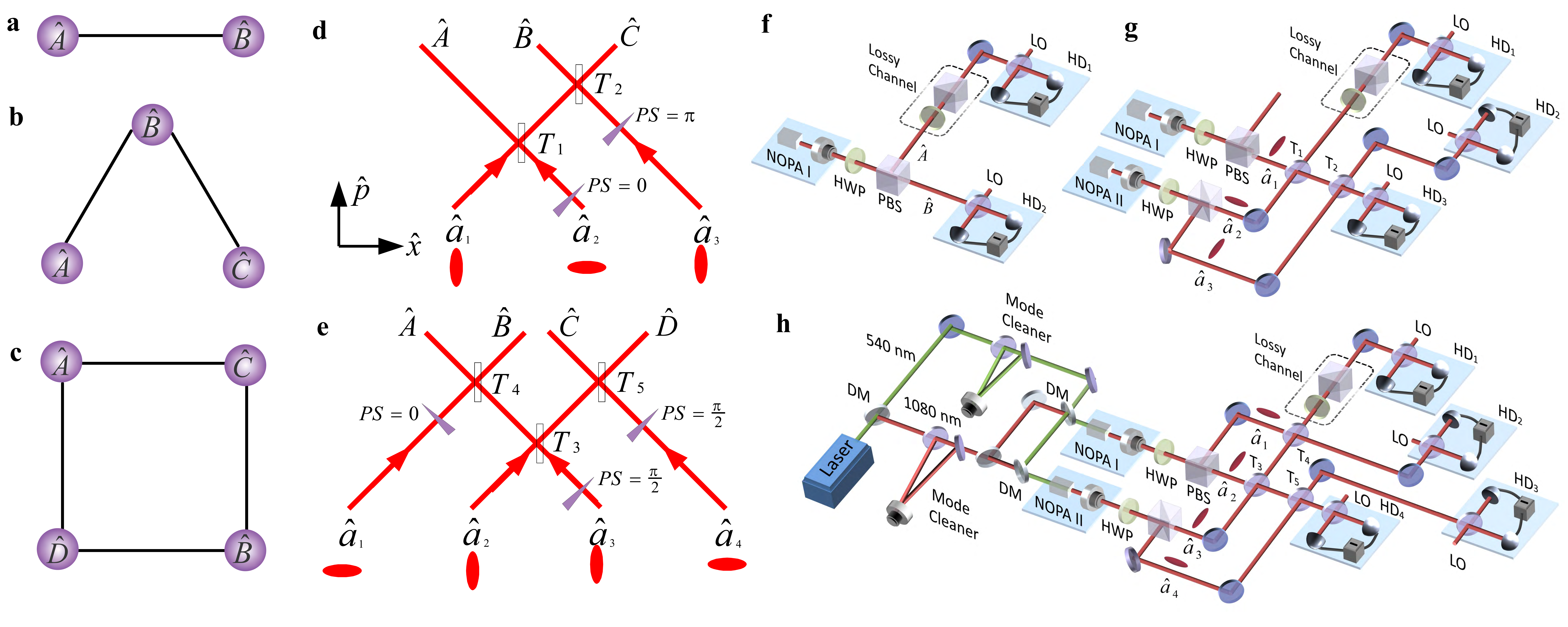}
\caption{Graph representation of multipartite CV entangled states and
their preparation. \textbf{a} CV two-mode Gaussian entangled state. \textbf{b}
Three-mode GHZ state. \textbf{c} Four-mode square Gaussian cluster state, respectively. \textbf{d} and \textbf{e} show the beam-splitter network used to generate the three-mode GHZ state and four-mode square Gaussian cluster
state, respectively. The phase shift (PS) is realized by locking the relative phase of two light beams at corresponding beam splitter. \textbf{f}, \textbf{g} and \textbf{h} show the schematics of
preparation and measuring the two-mode Gaussian entangled state, three-mode GHZ
state, and four-mode square Gaussian cluster state, respectively. PS, phase shift; NOPA, nondegenerate optical parametric amplifier; HWP, half-wave plate; PBS, polarizing beam splitter; LO, local oscillator; HD$_{1-4}$, homodyne detectors; DM, dichroic mirror.}
\label{fig:scheme}
\end{figure*}

Here, we analyze experimentally generated CV multi-mode entangled states of two, three and four photonic modes using the recently developed bosonic squeezing coefficients and the CV quantum Fisher information. Our complete microscopic mode-by-mode study encompasses all possible multi-partitions of the systems as well as the reduced distributions obtained by tracing over certain modes. A controllable loss channel on one of the modes is used to study the effect of losses on the multipartite entanglement structure. Our analysis is based on experimentally extracted covariance matrices and demonstrates the applicability of  entanglement criteria based on the Fisher information to CV systems and cluster states. Sudden transitions as a function of loss and noise-independent partitions are explained intuitively by the geometric interpretation of our entanglement criteria in phase space. Finally, we show that the criteria are not equivalent to the Gaussian PPT criterion, which can only be applied to bi-partitions.

\section*{Results}

\noindent \textbf{CV entanglement criteria from squeezing coefficients and Fisher information}

We consider an $N$-mode continuous-variable system with a vector of phase-space operators $\hat{\mathbf{r}}=(\hat{r}_1,\dots,\hat{r}_{2N})=(\hat{x}_1,\hat{p}_1,\dots,\hat{x}_N,\hat{p}_N)$. Any real vector $\mathbf{g}=(g_1,\dots,g_{2N})$ defines a multi-mode quadrature $\hat{q}(\mathbf{g})=\mathbf{g}\cdot\hat{\mathbf{r}}$, which generates displacements of the form $\hat{D}(\theta)=\exp(-i\hat{q}(\mathbf{g})\theta)$. The sensitivity of a Gaussian quantum state $\hat{\rho}$ under such displacements is determined by the quantum Fisher information \cite{BC94,Pinel13,Zhang2014}
\begin{align}\label{eq:QFI}
F_Q[\hat{\rho},\hat{q}(\mathbf{g})]=\mathbf{g}^T\boldsymbol{\Omega}^T\boldsymbol{\Gamma}^{-1}_{\hat{\rho}}\boldsymbol{\Omega}\mathbf{g},
\end{align}
where $\boldsymbol{\Omega}=\bigoplus_{i=1}^N\left(\begin{smallmatrix} 0 & 1\\-1 & 0 \end{smallmatrix}\right)$ is the symplectic form and $\boldsymbol{\Gamma}^{-1}_{\hat{\rho}}$ is the inverse of the covariance matrix with elements $(\boldsymbol{\Gamma}_{\hat{\rho}})_{ij}=\frac{1}{2}\langle\hat{r}_{i}\hat{r}_{j}+\hat{r}_{j}\hat{r}_{i}\rangle_{\hat{\rho}}-\langle\hat{r}_{i}\rangle_{\hat{\rho}}\langle\hat{r}_{j}\rangle_{\hat{\rho}}$. By means of the quantum Cram\'er-Rao inequality, the quantum Fisher information directly determines the precision bound for a quantum parameter estimation of $\theta$. It was shown in Ref.~\cite{GessnerPRA2016} that an upper limit for the sensitivity of mode-separable states is given in terms of the single-mode variances of the same state:
\begin{align}\label{eq:QFIsep}
F_Q[\hat{\rho}_{\mathrm{sep}},\hat{q}(\mathbf{g})]\leq 4\mathbf{g}^T\boldsymbol{\Gamma}_{\Pi(\hat{\rho}_{\mathrm{sep}})}\mathbf{g},
\end{align}
where $\boldsymbol{\Gamma}_{\Pi(\hat{\rho}_{\mathrm{sep}})}$ is the covariance matrix after all elements except the central $2\times 2$ blocks have been set to zero, effectively removing all mode correlations. This corresponds to the covariance matrix of the product state of the reduced density matrices $\Pi(\hat{\rho})=\bigotimes_{i=1}^N\hat{\rho}_i$. Any violation of inequality~(\ref{eq:QFIsep}) indicates the presence of entanglement between the modes. To identify the contribution of specific subsystems in a multipartite system, this criterion can be generalized for a microscopic analysis of the entanglement structure \cite{GessnerQuantum}. A witness for entanglement in a specific partition of the full system into subsystems $\Lambda=\mathcal{A}_1|\dots|\mathcal{A}_M$, where $\mathcal{A}_l$ describes an ensemble of modes, is obtained from Eq.~(\ref{eq:QFIsep}) by replacing the fully separable product state $\Pi(\hat{\rho})$ on the right-hand side by a product state on the partition $\mathcal{A}_1|\dots|\mathcal{A}_M$. More precisely, any $\mathcal{A}_1|\dots|\mathcal{A}_M$-separable quantum state, i.e., any state that can be written as $\hat{\rho}_{\Lambda}=\sum_{\gamma}p_{\gamma}\hat{\rho}^{(\gamma)}_{\mathcal{A}_1}\otimes\dots\otimes\hat{\rho}^{(\gamma)}_{\mathcal{A}_M}$, where $p_{\gamma}$ is a probability distribution, must satisfy \cite{GessnerQuantum}
\begin{align}\label{eq:QFIpsep}
F_Q[\hat{\rho}_{\Lambda},\hat{q}(\mathbf{g})]\leq 4\mathbf{g}^T\boldsymbol{\Gamma}_{\Pi_{\Lambda}(\hat{\rho}_{\Lambda})}\mathbf{g},
\end{align}
where $\Pi_{\Lambda}(\hat{\rho}_{\Lambda})=\bigotimes_{l=1}^M\hat{\rho}_{\mathcal{A}_l}$ and $\hat{\rho}_{\mathcal{A}_l}$ is the reduced density matrix of $\hat{\rho}_{\Lambda}$ on $\mathcal{A}_l$. The covariance matrix $\boldsymbol{\Gamma}_{\Pi_{\Lambda}(\hat{\rho}_{\Lambda})}$ can be easily obtained from $\boldsymbol{\Gamma}_{\hat{\rho}_{\Lambda}}$ by setting only those off-diagonal blocks to zero which describe correlations between different subsystems $\mathcal{A}_l$. The fully separable case, Eq.~(\ref{eq:QFIsep}), is recovered if each $\mathcal{A}_l$ contains exactly one mode.

\begin{figure}[tp]
\includegraphics[width=.49\textwidth]{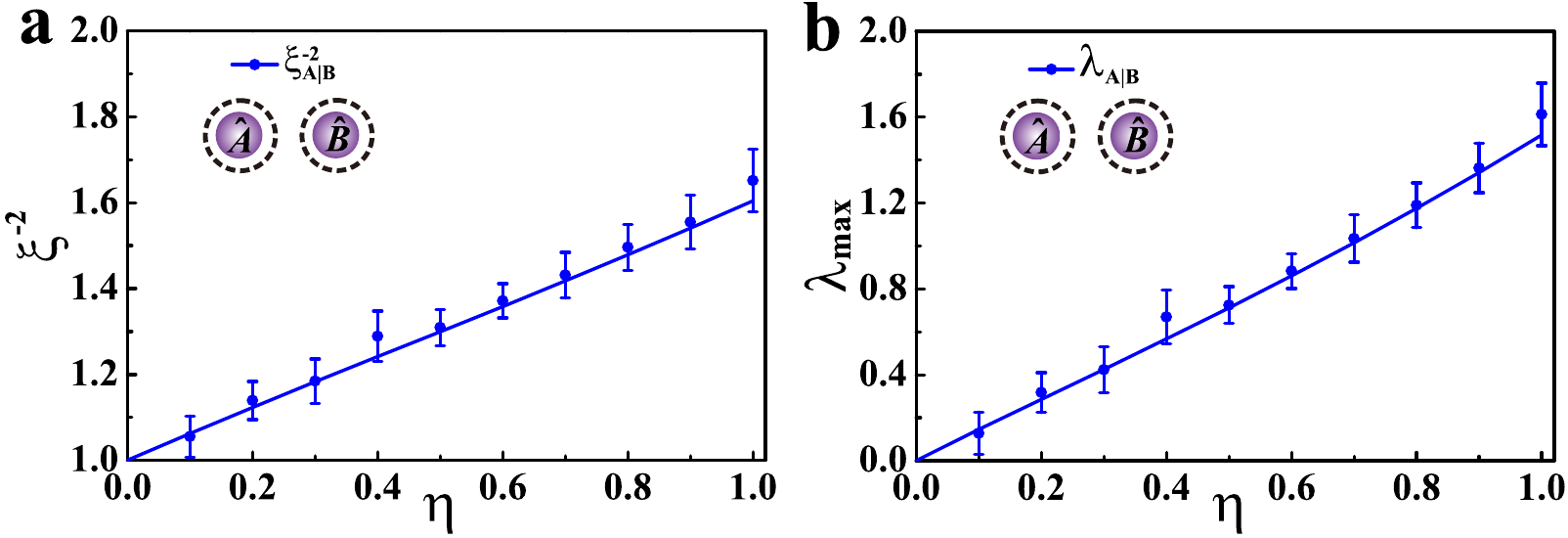}
\caption{Experimental results for the CV two-mode Gaussian entangled state in a lossy channel with transmission efficiency $\eta$. \textbf{a} Inverse multi-mode squeezing coefficients~(\ref{eq:chi2mult}). The plot shows the squeezing coefficient $\xi^{-2}_{A|B}$ obtained by numerically minimizing in Eq.~(\ref{eq:chi2mult}), using the experimentally measured covariance matrices (blue dots) and the theoretical prediction based on the state preparation schemes described in Fig.~\ref{fig:scheme} (blue line). Values above one violate~(\ref{eq:simplecovcrit2}) and indicate entanglement. \textbf{b} Gaussian quantum Fisher information entanglement criterion, expressed by the maximum eigenvalue of the matrix on the left-hand side (l.h.s.) of Eq.~(\ref{eq:Diffsep}). Positive values violate the separability condition~(\ref{eq:Diffsep}). The error bars represent one standard deviation and are obtained from the statistics of the measured data.}
\label{fig:EPR}
\end{figure}

By combining the separability criterion~(\ref{eq:QFIpsep}) with the expression for the quantum Fisher information of Gaussian states~(\ref{eq:QFI}), we find the following condition for the covariance matrix of $\mathcal{A}_1|\dots|\mathcal{A}_M$-separable states:
\begin{align}\label{eq:Diffsep}
\boldsymbol{\Gamma}^{-1}_{\hat{\rho}}-4\boldsymbol{\Omega}^T\boldsymbol{\Gamma}_{\Pi_{\Lambda}(\hat{\rho})}\boldsymbol{\Omega}\leq 0,
\end{align}
where we have used that both expressions~(\ref{eq:QFI}) and~(\ref{eq:QFIpsep}) are valid for arbitrary $\mathbf{g}$ and then multiplied the equation with $\boldsymbol{\Omega}$ from both sides using $\boldsymbol{\Omega}^T\boldsymbol{\Omega}=\mathbb{I}_{2N}$ and $\boldsymbol{\Omega}^T=-\boldsymbol{\Omega}$. Inequality~(\ref{eq:Diffsep}) expresses that the matrix on the left-hand side must be negative semidefinite. Hence, if we find a single positive eigenvalue, entanglement in the considered partition is revealed. Thus, it suffices to check whether the maximal eigenvalue $\lambda_{\max}$ is positive. The corresponding eigenvector $\mathbf{e}_{\max}$ further identifies a $2N$-dimensional ``direction'' in phase space such that the sensitivity under displacements generated by $\hat{q}(\mathbf{e}_{\max})$ maximally violates Eq.~(\ref{eq:QFIpsep}).

A lower bound on the quantum Fisher information of arbitrary states can be found from elements of the covariance matrix using \cite{GessnerQuantum}
\begin{align}\label{eq:lowerbound}
F_Q[\hat{\rho},\hat{q}(\mathbf{g})]\geq \frac{(\mathbf{h}^T\boldsymbol{\Omega}\mathbf{g})^2}{\mathbf{h}^T\boldsymbol{\Gamma}_{\hat{\rho}}\mathbf{h}},
\end{align}
which holds for arbitrary $\mathbf{g}$, $\mathbf{h}$. Choosing $\mathbf{h}=\boldsymbol{\Omega}\mathbf{g}$ with $|\mathbf{g}|^2=1$ leads with~(\ref{eq:QFIpsep}) to the separability condition \cite{GessnerQuantum}
\begin{align}\label{eq:simplecovcrit2}
\xi_{\Lambda}^{-2}(\hat{\rho}_{\mathrm{sep}})\leq1,
\end{align}
where
\begin{align}\label{eq:chi2mult}
\xi_{\Lambda}^2(\hat{\rho}):=\min_{\mathbf{g}}4(\mathbf{g}^T\boldsymbol{\Omega}^T\boldsymbol{\Gamma}_{\Pi_{\Lambda}(\hat{\rho})}\boldsymbol{\Omega}\mathbf{g})(\mathbf{g}^T\boldsymbol{\Gamma}_{\hat{\rho}}\mathbf{g}),
\end{align}
is the bosonic multi-mode squeezing coefficient for the partition $\Lambda$. Here, the minimizing $\mathbf{g}$ can be interpreted as a direction in phase space that identifies a multi-mode quadrature $\hat{q}(\mathbf{g})$ with a squeezed variance which can be traced back to mode entanglement \cite{GessnerQuantum}.

\begin{figure}[tp]
\includegraphics[width=.49\textwidth]{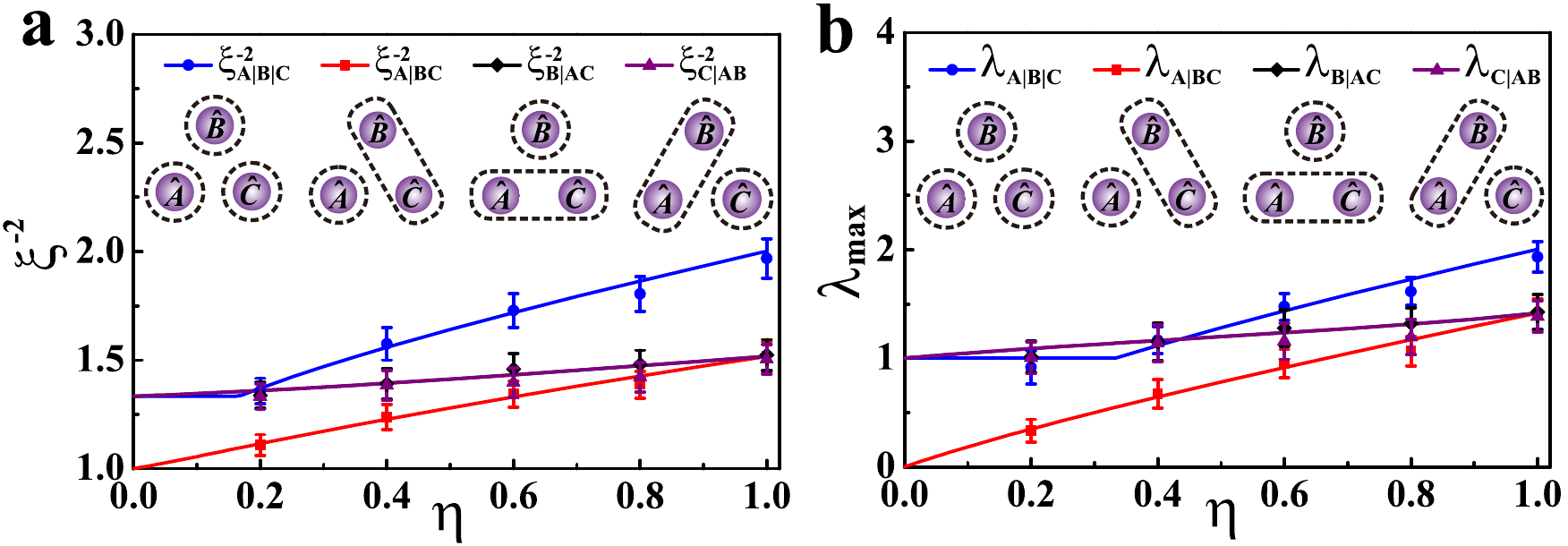}
\caption{Experimental results for the three-mode GHZ state in a lossy channel with transmission efficiency $\eta$. \textbf{a} Inverse multi-mode squeezing coefficients. \textbf{b} Gaussian Fisher information entanglement criterion. Shown are numerically optimized coefficients for the partitions $A|B|C$ (blue circles), $A|BC$ (red squares), $B|AC$ (black diamonds), and $C|AB$ (purple triangles) from experimentally obtained covariance matrices and the curves represent the theoretical prediction.}
\label{fig:GHZ}
\end{figure}

\noindent \textbf{Experimental setup}

In the following we analyze experimentally generated $N$-mode Gaussian states with $N=2,3,4$, subject to asymmetric loss using the two entanglement criteria defined by the quantum Fisher information, Eq.~(\ref{eq:Diffsep}), and the multi-mode squeezing coefficient, Eq.~(\ref{eq:simplecovcrit2}). The graph representations of the three classes of Gaussian multi-mode entangled states considered here are shown in Fig. 1. They are often referred to as CV two-mode Gaussian entangled state ($N=2$, Fig. 1a), three-mode CV Greenberger-Horne-Zeilinger (GHZ) state ($N=3$, Fig. 1b), and four-mode square Gaussian cluster state ($N=4$, Fig. 1c). The experimental generation of the states is described in detail in Materials and Methods and Refs.~\cite{SciRep2017,PRL2017}. In all cases, the CV entangled states are generated by nondegenerate optical parametric amplifiers (NOPAs) with $-3$ dB squeezing at the sideband frequency of $3$ MHz. The two-mode Gaussian entangled state is prepared directly by a NOPA. The three-mode GHZ state is obtained by combining a phase-squeezed and two amplitude-squeezed states using two beam splitters with transmissivities of $T_{1}=1/3$ and $T_{2}=1/2$, respectively, as shown in Fig. 1d \cite{SciRep2017}. Similarly, the four-mode square Gaussian cluster state is prepared by coupling two phase-squeezed and two amplitude-squeezed states on a beam-splitter network consisting of three beam splitters with $T_{3}=1/5$ and $T_{4}=T_{5}=1/2$, respectively, as shown in Fig. 1e \cite{PRL2017}.

To study the robustness of multipartite entanglement under transmission losses, a lossy quantum channel for mode $A$ is simulated using a half-wave plate (HWP) and a polarizing beam splitter (PBS). The output mode is given by $\hat{A}^{\prime }=\sqrt{\eta }%
\hat{A}+\sqrt{1-\eta }\hat{\upsilon}$, where $\eta $ and $\hat{\upsilon}$ represent the transmission efficiency of the quantum channel and the vacuum mode induced by loss into the quantum channel, respectively, as shown in Fig. 1f--h. Let us now turn to the characterization of CV entanglement based on the experimentally generated data.

\begin{figure*}[tp]
\includegraphics[width=.98\textwidth]{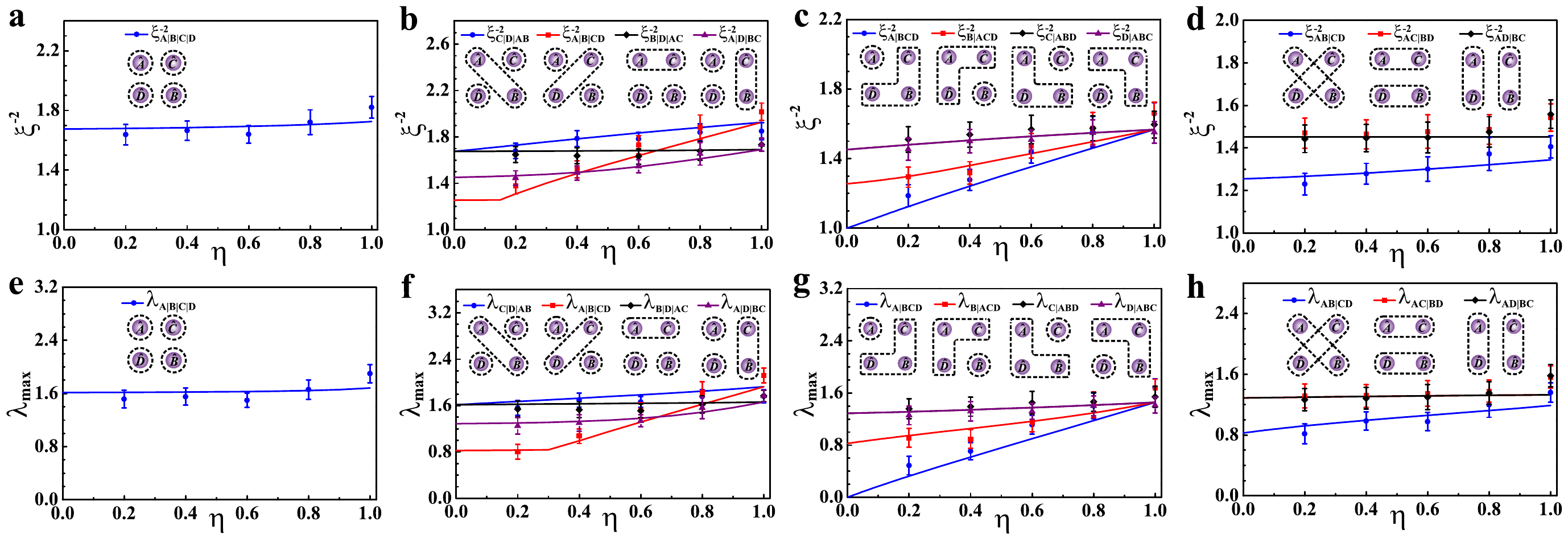}
\caption{Experimental results for the four-mode square Gaussian cluster state in a lossy channel with transmission efficiency $\eta$. \textbf{a--d} Inverse multi-mode squeezing coefficients $\xi^{-2}$ for the partitions of classes $1\otimes1\otimes1\otimes1$, $1\otimes1\otimes2$, $1\otimes3$, and $2\otimes2$, respectively. \textbf{e--h} The corresponding data for Gaussian Fisher information entanglement criterion. The data points are numerically optimized coefficients from experimentally obtained covariance matrices and the curves represent the corresponding numerically optimized predictions from the theoretical model.}
\label{fig:Cluster}
\end{figure*}

\noindent \textbf{Experimental results} \label{sec:epr}

Fig. 2a shows the inverse squeezing coefficient~(\ref{eq:chi2mult}) $\xi^{-2}_{A|B}$ for an CV two-mode Gaussian entangled state in a lossy channel for the only possible partition $A|B$ of the bipartite system. The coefficient $\xi^{-2}_{A|B}$ decreases as the transmission efficiency $\eta$ decreases but it always violates the separability condition~(\ref{eq:simplecovcrit2}) unless $\eta=0$, i.e., when mode $A$ is completely lost. This confirms that CV  two-mode Gaussian entanglement only decreases but never fully disappears due to particle losses, i.e., CV  two-mode Gaussian entanglement is robust to loss \cite{Robustness}. We observe the same behavior for the criterion Eq.~(\ref{eq:Diffsep}), which makes use of the Gaussian quantum Fisher information. Fig. 2b shows the maximum eigenvalue $\lambda_{\max}$ of the matrix $\boldsymbol{\Gamma}^{-1}_{\hat{\rho}}-4\boldsymbol{\Omega}^T\boldsymbol{\Gamma}_{\hat{\rho}_A\otimes\hat{\rho}_B}\boldsymbol{\Omega}$. According to Eq.~(\ref{eq:Diffsep}), a positive value indicates entanglement. Both coefficients attain their two-fold degenerate maximal value for the phase space directions $\mathbf{g}=(\sin\phi,0,\cos\phi,0)$ and $\mathbf{g}=(0,-\sin\phi,0,\cos\phi)$, where $\phi$ is a function of $\eta$ (for $\eta=1$ we have $\phi=\pi/4$ \cite{GessnerQuantum}). These directions indicate strong correlations in the momentum quadratures and anti-correlations in the position quadratures, allowing us to relate the entanglement to the squeezing of the collective variances $\Delta(\hat{x}_A\sin\phi+\hat{x}_B\cos\phi)^2$ and $\Delta(\hat{p}_A\sin\phi-\hat{p}_B\cos\phi)^2$. It should be noted that $\xi^{-2}_{A|B}$ and $\xi^{-2}_{B|A}$ ($\lambda_{A|B}$ and $\lambda_{B|A}$) are identical because the entanglement coefficients only depend on the partition and not on the order in which the subsystems are labeled.

The entanglement structure becomes more interesting for the
three-mode GHZ state, exhibiting four non-trivial partitions of the
system, as well as three reduced two-mode states. The squeezing
coefficient~(\ref{eq:chi2mult}), as well as the Gaussian Fisher
information entanglement criterion~(\ref{eq:Diffsep}), are plotted
in Fig. 3 for all four partitions. Both show that at
$\eta=1$ the three bi-separable partitions $A|BC$, $B|AC$, and
$C|AB$ are equivalent due to the symmetry of the state, but as
$\eta$ is decreased, the entanglement in the partition $A|BC$ is
more strongly affected by the losses than that of the other two
partitions. In the extreme case where mode $A$ is fully lost
($\eta=0$) there is still some residual entanglement between $B$ and
$C$ \cite{AdessoThreeModes}. In this case, all partitions are
equivalent to the bi-partition $B|C$. The data shown in
Fig. 3 confirms this: In both cases, the entanglement
witness for all partitions coincide at $\eta=0$, except $A|BC$
which, as expected, yields zero.

We further notice a discontinuity for the theoretical predictions of both witnesses regarding the fully separable partition $A|B|C$ as a function of $\eta$ (blue lines in Fig. 3). This can be explained by analyzing the corresponding optimal phase space direction $\mathbf{g}$. In the presence of only moderate losses, the maximal correlations and squeezing are identified along the direction $\mathbf{g}=(0,c_1,0,c_2,0,c_2)$ with $c_1^2+2c_2^2=1$, i.e., the multi-mode quadrature $\hat{q}(\mathbf{g})=c_1\hat{p}_A+c_2\hat{p}_B+c_2\hat{p}_C$ which involves all three modes. The squeezing along this phase-space direction diminishes with increasing losses. When the losses of mode $A$ become dominant, the squeezing along the phase space direction $\mathbf{g}=(0,0,1,0,-1,0)/\sqrt{2}$, i.e., of the quadrature $\hat{q}(\mathbf{g})=(\hat{x}_B-\hat{x}_C)/\sqrt{2}$ is more pronounced as it does not decay with $\eta$, being independent of mode $A$. The discontinuity is therefore explained by a sudden change of the optimal squeezing direction due to depletion of mode $A$. We remark that the experimentally prepared states are the same, except for the variable $\eta$. The change of the squeezing direction simply implies that when the local noise exceeds a critical value, the entanglement is more easily revealed by analyzing the quantum state from a different `perspective' in phase space. Notice that having access to the full covariance matrix, we can analyze both entanglement witnesses for arbitrary directions.

The change of the optimal direction is observed for both
entanglement coefficients, whereas the transition occurs at a larger
value of $\eta$ for the Fisher information
criterion~(\ref{eq:Diffsep}) (see Supplementary Information). There we also show the two-mode
entanglement properties after tracing over one of the modes in an
analysis of the reduced density matrices, which show that two-mode
entanglement persists after tracing over one of the subsystems, in
stark contrast to GHZ states of discrete variables \cite{Dur}.

Finally, we analyze the four-mode square Gaussian cluster state in Fig. 4. We find that the decoherence of entanglement depends on the cluster state's geometric structure. As shown in Fig. 4a, the inverse multi-mode squeezing coefficient $\xi^{-2}_{A|B|C|D}$ for the fully separable partition is not sensitive to transmission loss on mode $A$, while decoherence affects the coefficients for other partitions shown in Fig. 4b--d. For $1\otimes1\otimes2$ partitions, only the results of $\xi^{-2}_{C|D|AB}$, $\xi^{-2}_{A|B|CD}$, $\xi^{-2}_{B|D|AC}$, and $\xi^{-2}_{A|D|BC}$ are shown in Fig. 4b ($\xi^{-2}_{B|C|AD}$ and $\xi^{-2}_{A|C|BD}$ are shown in Fig. S3 in  Supplementary Information). The discontinuity for the $A|B|CD$ partition is again explained by a transition of the optimal squeezing direction at a critical value of the transmission $\eta$ for the isolated mode $A$ (see Supplementary Information). The two coefficients $\xi^{-2}_{C|D|AB}$ and $\xi^{-2}_{A|B|CD}$ ($\xi^{-2}_{B|D|AC}$ and $\xi^{-2}_{A|D|BC}$) are equal for $\eta=1$ because of the symmetric roles of these modes in these partitions. As shown in Fig. 4b and Fig. 4c, the most sensitive coefficients to transmission losses of mode $A$ are those where mode $A$ is an individual subsystem. The coefficients $\xi^{-2}_{C|ABD}$ and $\xi^{-2}_{D|ABC}$ overlap due to the symmetric roles of modes $C$ and $D$.

Fig. 4d shows the inverse multi-mode squeezing coefficients for $2\otimes2$ partitions. It is interesting that the coefficient $\xi^{-2}_{AC|BD}$ ($\xi^{-2}_{AD|BC}$) is immune to transmission loss of mode $A$. This indicates that the collective coefficients for $2\otimes2$ partitions, where each partition is composed by two neighboring modes (recall the graph representation in Fig. 1c), is not sensitive to the loss of any one mode. In contrast, the coefficient $\xi^{-2}_{AB|CD}$, where each subsystem is composed by two diagonal modes, is still sensitive to transmission loss. As before, we find that the qualitative behavior of the squeezing coefficient $\xi^{-2}$ coincides with that of $\lambda_{\max}$ of the Gaussian Fisher information criterion~(\ref{eq:Diffsep}), see Fig. 4e--h.

A further understanding of the entanglement structure is provided by an analysis of the three-mode and two-mode reduced density matrices of the state as well as of the optimal directions. A detailed analysis reveals that the loss-robustness is drastically reduced for all partitions if either mode $C$ or $D$ is traced out (see Supplementary Information). Moreover, for very small values of $\eta$, the entanglement in the partitions $A|CD$, $D|AB$ and $C|AB$ in the reduced three-mode states is revealed by the criterion~(\ref{eq:Diffsep}) but not by the squeezing approximation~(\ref{eq:chi2mult}), where we assumed $\mathbf{h}=\boldsymbol{\Omega}\mathbf{g}$ to simplify the optimization (see Supplementary Information).

\section*{Discussion}
To benchmark our CV entanglement criteria, we may compare them to the PPT criterion, which is necessary and sufficient for $1\otimes (N-1)$ separability of Gaussian states \cite{PPTCriteria,DuanCriteria,WernerWolf}. For partitions into more than two subsystems the PPT criterion cannot be applied. It is interesting to notice that the PPT separability condition can be expressed as
\begin{align}\label{eq:PPT}
\boldsymbol{\Gamma}_{\hat{\rho}^{\mathrm{PPT}}}^{-1}-4\boldsymbol{\Omega}^T\boldsymbol{\Gamma}_{\hat{\rho}^{\mathrm{PPT}}}\boldsymbol{\Omega}\leq 0,
\end{align}
where $\Gamma_{\hat{\rho}^{\mathrm{PPT}}}$ is the covariance matrix after application of the partial transposition operation. The condition~(\ref{eq:PPT}) is the Heisenberg-Robertson uncertainty relation for the state $\hat{\rho}^{\mathrm{PPT}}$ and constitutes a \textit{bona-fide} condition for the physicality of the covariance matrix \cite{Simon1994}. A violation of~(\ref{eq:PPT}) therefore indicates that $\Gamma_{\hat{\rho}^{\mathrm{PPT}}}$ does not correspond to a physical state, from which one can conclude that the original state, described by $\Gamma_{\hat{\rho}}$, is entangled. The similarity of Eqs.~(\ref{eq:Diffsep}) and~(\ref{eq:PPT}) indicates the close relationship of the two criteria with the uncertainty relation \cite{GessnerQuantum}. Note also that all pure Gaussian states $\Psi$ saturate the uncertainty relation $\boldsymbol{\Gamma}_{\Psi}^{-1}=4\boldsymbol{\Omega}^T\boldsymbol{\Gamma}_{\Psi}\boldsymbol{\Omega}$, which together with Eq.~(\ref{eq:QFI}) allows us to recover the relation $F_Q[\Psi,\hat{q}(\mathbf{g})]=4\mathbf{g}^T\boldsymbol{\Gamma}_{\Psi}\mathbf{g}$ for pure states.

The criteria employed here and the well-established PPT condition are not equivalent. This is indicated by: (i) the applicability of our criteria to multi-partite separability classes, their geometric interpretation and their connection to metrological sensitivity. For example, we analyzed a total of 44 partitions for CV multi-mode entangled states of two, three and four photonic modes. Out of these, 12 are genuine multi-partitions, which can not be analyzed with the PPT criterion (see Supplementary Information for more details). (ii) the ability to detect non-Gaussian entanglement beyond the PPT condition using the CV Fisher information \cite{GessnerPRA2016} or squeezing of higher-order observables \cite{GessnerQuantum}. (iii) the existence of PPT-entangled states which are not revealed by the CV squeezing coefficient or the Fisher information for displacements. A simple class of states that belong to (iii) can be constructed by mixing the two-mode Gaussian entangled state with the vacuum (see Supplementary Information). Moreover, the reduced two-mode states of the four-mode cluster state studied here also are examples of (iii).

In conclusion, we demonstrated that the multi-mode squeezing coefficient and the quantum Fisher information provide useful tools to understand the entanglement structure of Gaussian $N$-mode entangled states. In our microscopic analysis of CV states of up to four modes we characterized the robustness of entanglement for each partition individually. The effect of losses on more than one mode of three-mode and four-mode entangled states are also theoretically investigated, which confirms the resilience of multipartite CV Gaussian entanglement to finite losses (see Supplementary Information). The methods employed in this work yield a geometric interpretation in terms of a phase-space direction that identifies a strongly squeezed multi-mode quadrature as the origin of the mode correlations. Certain partitions revealed sudden transitions of the optimal phase-space direction for entanglement detection, rendering the entanglement coefficient invariant after passing a threshold value. This is strongly reminiscent of the ``freezing" behavior previously observed for measures of entanglement \cite{Carnio1}, discord \cite{Haikka} and coherence \cite{Bromley} under incoherent dynamics. However, it is important to notice that the entanglement criteria considered here are witnesses of entanglement and do not represent quantitative measures in a strict sense.

The squeezing coefficient represents an easily accessible entanglement criterion, based on a second-order approximation of the quantum Fisher information, which is more involved to extract experimentally for general states. For the specific case of Gaussian states, both criteria are expressed in terms of moments up to second order, but for the squeezing coefficient the optimization was restricted to specific quadratures to reduce the number of parameters. This was found to be a suitable approximation in most cases, as we obtained qualitatively equivalent results to the Fisher information. Only in the presence of strong losses, the Fisher information reveals Gaussian entanglement for certain partitions of the reduced states that remains undisclosed by the squeezing coefficient.

Our detailed analysis highlights the advantages of the mode entanglement criteria based on the quantum Fisher information for Gaussian states, in particular, their ability to study multi-partitions based on available data only, their geometric interpretation, and their relation to the metrological sensitivity. We have also observed their limitations, i.e., not being a necessary and sufficient condition for all Gaussian states. However, in principle the entanglement of arbitrary pure states can be revealed using the Fisher information criterion \cite{GessnerPRA2016}. These methods thus complement the well-established PPT techniques for CV systems, which are necessary and sufficient for Gaussian $1\otimes (N-1)$ systems but unfitting for multi-partitions and of limited applicability for non-Gaussian states.

The more general criterion based on the quantum Fisher information is expected to be particularly useful for non-Gaussian states. In this case, it is able to reveal entanglement even when entanglement criteria based on second-order moments can no longer be applied and the concept of squeezing is ill-defined. We expect that these methods provide useful techniques for the analysis of entanglement in complex CV networks \cite{Seiji}.

\section*{Methods}
\noindent \textbf{Details of experiment}

The experimental setup to generate the CV two-mode Gaussian entangled state is depicted in Fig. 1f. A $-3$ dB two-mode Gaussian entangled state at the sideband frequency of $3$ MHz is generated directly
from a NOPA I. The lossy channel (LC) is composed by a half-wave
plate (HWP) and a polarization beam splitter (PBS). Quadratures are
measured via homodyne detectors (HD)$_{1-2}$ and the local
oscillator (LO).

The three-mode GHZ state is generated using the experimental setup depicted in Fig. 1g. The squeezed states are generated from the coupled modes at $+45^{\circ }$ and $-45^{\circ }$ polarization directions of two NOPAs. Further technical details can be found in Ref.~\cite{SciRep2017}.

Fig. 1h depicts the experimental setup used to generate the four-mode Gaussian square cluster state. A dual-wavelength laser for 540 nm and 1080 nm is used. Two mode cleaners are inserted between the
laser source and the NOPAs to filter higher order spatial modes and noise of the laser beams at two wavelengths, respectively. In addition to elements described already for Fig. 1f and Fig. 1g, dichroic mirrors (DMs) are also shown. For technical details we refer to Ref.~\cite{PRL2017}.

\noindent \textbf{Reconstruction of covariance matrices}

In the experiment, the covariance matrices of the multipartite CV entangled states are obtained from local measurements on the optical output modes. These measurements include the amplitude and phase quadratures $\Delta ^{2}\hat{r}_{i}$, $\Delta ^{2}\hat{r}_{j}$, and the cross correlations $\Delta
^{2}\left( \hat{r}_{i}+\hat{r}_{j}\right) $ or $\Delta
^{2}\left( \hat{r}_{i}-\hat{r}_{j}\right) $.
The elements of the covariance matrix are calculated via the identity
\begin{align}
(\boldsymbol{\Gamma}_{\hat{\rho}})_{ij} & =\frac{1}{2}\left[ \Delta
^{2}\left( \hat{r}_{i}+\hat{r}_{j}\right) -\Delta ^{2}\hat{r}%
_{i}-\Delta ^{2}\hat{r}_{j}\right] ,  \notag \\
(\boldsymbol{\Gamma}_{\hat{\rho}})_{ij} & =-\frac{1}{2}\left[ \Delta
^{2}\left( \hat{r}_{i}-\hat{r}_{j}\right) -\Delta ^{2}\hat{r}%
_{i}-\Delta ^{2}\hat{r}_{j}\right] .
\end{align}

For each transmission efficiency $\eta$ of mode $A$, three sets of covariance matrices are reconstructed. Error bars for all the experimental data are obtained from the
statistics of the three covariance matrices.

\textbf{Data availability.} The data that support the findings of this study are available from the corresponding author on request.

\textbf{Acknowledgments.}
This research was supported by the National Natural Science Foundation of China (Grants No. 11522433, NO. 11834010, No. 61601270, No. 61475092 and No. 11874247), National Key R\&D Program of China (Grants No.
2016YFA0301402, No. 2017YFA0304500, and No.2017YFA0304203), 111 project (Grant No. D18001), the Hundred Talent Program of the Shanxi Province (2018), Fund for Shanxi ``1331" Project Key Subjects Construction, and the European Commission through the Quant-ERA
project ``CEBBEC". X. S. thanks the program of Youth Sanjin Scholar. M. G. thanks the Alexander von Humboldt foundation for support. W. L. thanks PCSIRT (Grant No. IRT\_17R70) and the Program of State Key Laboratory of Quantum Optics and Quantum Optics Devices (No. KF201703).

\textbf{Author contributions.}
M. G., W. L., X. S. and A. S. conceived the idea. X. S. and K. P. designed the experiment. Z. Q., X. D. and X. S. constructed and performed the experiment. D. H. participated in part of the experiment. Z. Q., Z. R. and M. G. analyzed the data. M. G. and Z. Q. wrote the manuscript with input from Z. R., W. L., X. S. and A. S. All authors participated in the discussion of the results and commented on the manuscript.

Z. Q., M. G. and Z. R. contributed equally to this work.

\textbf{Competing interests}: The authors declare no competing financial or non-financial conflicts.

\clearpage

\begin{center}
{\large \bfseries Supplementary Material}
\end{center}

\setcounter{equation}{0} \setcounter{figure}{0} \renewcommand\thefigure{S%
\arabic{figure}} \renewcommand\theequation{S\arabic{equation}}

\section{Theoretical model}
The effect of the beam-splitter array on the initial product state of $N=3$ or $N=4$ modes, as depicted in Fig.1e and Fig.1f in the main text can be analytically predicted as a function of the squeezing strength $r$ and the transmission efficiency $\eta$ for mode $A$. Assuming the transmissivity coefficients $T_1,\dots,T_5$ as stated in the main text, we obtain the covariance matrices
\begin{align}
\boldsymbol{\Gamma}^{(2)}_{r,\eta}=\frac{1}{2}\left(
\begin{array}{cccc}
(1-\eta )+\eta c & 0 & -\sqrt{\eta } s & 0 \\
 0 & (1-\eta )+\eta c & 0 & \sqrt{\eta } s \\
 -\sqrt{\eta } s & 0 &   c & 0 \\
 0 & \sqrt{\eta } s & 0 &   c \\
\end{array}
\right),
\end{align}
for the two-mode states, 
\begin{widetext}
\begin{align}
\boldsymbol{\Gamma}^{(3)}_{r,\eta}=\left(
\begin{array}{cccccc}
 \frac{1}{6} \left(\left(-3+2 e^{-2 r}+e^{2 r}\right) \eta +3\right) & 0 & \frac{s \sqrt{\eta }}{3} & 0 & \frac{s \sqrt{\eta }}{3} & 0 \\
 0 & \frac{1}{6} (3 c \eta +s \eta -3 \eta +3) & 0 & -\frac{s \sqrt{\eta }}{3} & 0 & -\frac{s \sqrt{\eta }}{3} \\
 \frac{s \sqrt{\eta }}{3} & 0 & \frac{1}{6} e^{-2 r} \left(2+e^{4 r}\right) & 0 & \frac{s}{3} & 0 \\
 0 & -\frac{s \sqrt{\eta }}{3} & 0 & \frac{1}{6} (3 c+s) & 0 & -\frac{s}{3} \\
 \frac{s \sqrt{\eta }}{3} & 0 & \frac{s}{3} & 0 & \frac{1}{6} e^{-2 r} \left(2+e^{4 r}\right) & 0 \\
 0 & -\frac{s \sqrt{\eta }}{3} & 0 & -\frac{s}{3} & 0 & \frac{1}{6} (3 c+s) \\
\end{array}
\right),
\end{align}
for the three-mode states, and
\begin{align}
\boldsymbol{\Gamma}^{(4)}_{r,\eta}=\left(
\begin{array}{cccccccc}
 \frac{1}{10} (5 c \eta +s \eta -5 \eta +5) & 0 & -\frac{2 s \sqrt{\eta }}{5} & 0 & 0 & \frac{s \sqrt{\eta }}{5} & 0 & \frac{s \sqrt{\eta }}{5} \\
 0 & \frac{1}{10} (5 c \eta -s \eta -5 \eta +5) & 0 & \frac{2 s \sqrt{\eta }}{5} & \frac{s \sqrt{\eta }}{5} & 0 & \frac{s \sqrt{\eta }}{5} & 0 \\
 -\frac{2 s \sqrt{\eta }}{5} & 0 & \frac{1}{10} (5 c+s) & 0 & 0 & \frac{s}{5} & 0 & \frac{s}{5} \\
 0 & \frac{2 s \sqrt{\eta }}{5} & 0 & \frac{1}{10} (5 c-s) & \frac{s}{5} & 0 & \frac{s}{5} & 0 \\
 0 & \frac{s \sqrt{\eta }}{5} & 0 & \frac{s}{5} & \frac{1}{10} (5 c+s) & 0 & -\frac{2 s}{5} & 0 \\
 \frac{s \sqrt{\eta }}{5} & 0 & \frac{s}{5} & 0 & 0 & \frac{1}{10} (5 c-s) & 0 & \frac{2 s}{5} \\
 0 & \frac{s \sqrt{\eta }}{5} & 0 & \frac{s}{5} & -\frac{2 s}{5} & 0 & \frac{1}{10} (5 c+s) & 0 \\
 \frac{s \sqrt{\eta }}{5} & 0 & \frac{s}{5} & 0 & 0 & \frac{2 s}{5} & 0 & \frac{1}{10} (5 c-s) \\
\end{array}
\right),
\end{align}
\end{widetext}
for the four-mode states, respectively. We have abbreviated the functions $c=\cosh (2r)$ and $s=\sinh(2r)$.

\section{Microscopic entanglement structure}

\begin{table*}[p]
\begin{tabular}[c]{|c|c|cc|cc|c|c|cc|}\hline
&& \multicolumn{2}{|c|}{sqz., Eq.~(7)} & \multicolumn{2}{|c|}{QFI, Eq.~(4)} & & \multicolumn{3}{|c|}{detected} \\
state & partition &  $\xi^{-2}$ & $\mathbf{g}_{\min}$ & $\lambda_{\max}$ & $\mathbf{e}_{\max}$ & disc. & QFI / sqz. & PPT &\\\hline
\textbf{EPR} & $A|B$ & Fig.~2 & Main text & Fig.~2 & Main text & no & yes & yes &\\\hline
\textbf{GHZ} & $A|B|C$ & Fig.~3 & Tab.~\ref{tab:GHZ} & Fig.~3 & Tab.~\ref{tab:GHZ} & Fig.~\ref{fig:AsBsC} & yes &  N.A. &\\\cline{2-10}
 & $A|BC$ & \multirow{3}{*}{Fig.~3} & \multirow{3}{*}{Tab.~\ref{tab:GHZ}}& \multirow{3}{*}{Fig.~3} & \multirow{3}{*}{Tab.~\ref{tab:GHZ}}&  no &yes &  yes &\\
 & $AB|C$ &  & &  &  & no & yes & yes &\\
 & $AC|B$ &  &  &  &  & no & yes & yes &\\\cline{2-10}
reduced & $A|B$ & \multicolumn{2}{c|}{\multirow{3}{*}{Fig.~\ref{fig:reduced3}}} & \multicolumn{2}{c|}{\multirow{3}{*}{Fig.~\ref{fig:reduced3}}} & no & yes &  yes &\\
2 modes & $A|C$ &  &  &  &  & no & yes & yes &\\
 & $B|C$ &  & &  & & no & yes & yes &\\\hline
\textbf{Cluster} & $A|B|C|D$ & Fig.~4 & Tab.~\ref{tab:1111} & Fig.~4 & Tab.~\ref{tab:1111} & no & yes & N.A. &\\\cline{2-10}
 & $A|B|CD$ & \multirow{4}{*}{Fig.~4} & \multirow{4}{*}{Tab.~\ref{tab:112}} & \multirow{4}{*}{Fig.~4} & \multirow{4}{*}{Tab.~\ref{tab:112}} & Fig.~\ref{fig:AsBsCD} & yes & N.A. &\\
 & $A|D|BC$ &  &  &  &  & no & yes & N.A. &\\
 & $B|D|AC$ &  &  &  &  & no & yes & N.A. &\\
 & $C|D|AB$ &  &  &  &  & no & yes & N.A. &\\
 & $A|C|BD$ & \multirow{2}{*}{Fig.~\ref{fig:Cluster_Supp}} & \multirow{2}{*}{Tab.~\ref{tab:112}} & \multirow{2}{*}{Fig.~\ref{fig:Cluster_Supp}} & \multirow{2}{*}{Tab.~\ref{tab:112}} & no & yes & N.A. &\\
 & $B|C|AD$ &  &  &  &  & no & yes & N.A. &\\\cline{2-10}
 & $A|BCD$ & \multirow{4}{*}{Fig.~4} & \multirow{4}{*}{Tab.~\ref{tab:13}} & \multirow{4}{*}{Fig.~4} & \multirow{4}{*}{Tab.~\ref{tab:13}} & no & yes & yes & \multirow{4}{*}{Fig.~\ref{fig:PPT}}\\
 & $B|ACD$ &  &  &  &  & no & yes & yes &\\
 & $C|ABD$ &  &  &  &  & no & yes & yes &\\
 & $D|ABC$ &  &  &  &  & no & yes & yes &\\\cline{2-10}
 & $AB|CD$ & \multirow{3}{*}{Fig.~4} & \multirow{3}{*}{Tab.~\ref{tab:22}} & \multirow{3}{*}{Fig.~4} & \multirow{3}{*}{Tab.~\ref{tab:22}} & no & yes & yes  & \multirow{3}{*}{Fig.~\ref{fig:PPT}}\\
 & $AC|BD$ &  &  &  &  & no & yes & yes &\\
 & $AD|BC$ &  &  &  &  & no & yes & yes &\\\cline{2-10}
reduced & $B|C|D$ & \multicolumn{2}{c|}{Fig.~\ref{fig:reduced43S}} & \multicolumn{2}{c|}{Fig.~\ref{fig:reduced43F}} & no & yes &  N.A. &\\\cline{2-10}
3 modes & $B|CD$ & \multicolumn{2}{c|}{\multirow{3}{*}{Fig.~\ref{fig:reduced43S}}} & \multicolumn{2}{c|}{\multirow{3}{*}{Fig.~\ref{fig:reduced43F}}}  & no & yes & yes &\\
 & $C|BD$ &   &   &   &  & no & yes & yes &\\
 & $D|BC$ &   &   &   &   & no & yes & yes &\\\cline{2-10}
 & $A|C|D$ & \multicolumn{2}{c|}{Fig.~\ref{fig:reduced43S}} & \multicolumn{2}{c|}{Fig.~\ref{fig:reduced43F}} & yes & yes & N.A. &\\\cline{2-10}
 & $A|CD$ & \multicolumn{2}{c|}{\multirow{3}{*}{Fig.~\ref{fig:reduced43S}}} & \multicolumn{2}{c|}{\multirow{3}{*}{Fig.~\ref{fig:reduced43F}}}   & no$^{*}$ & yes / $\eta>0.13$ & yes &\\
 & $C|AD$ &   &   &   &   &  no & yes & yes &\\
 & $D|AC$ &   &   &   &   &  no & yes & yes &\\\cline{2-10}
 & $A|B|D$ & \multicolumn{2}{c|}{Fig.~\ref{fig:reduced43S}} & \multicolumn{2}{c|}{Fig.~\ref{fig:reduced43F}} & no & yes & N.A. &\\\cline{2-10}
 & $A|BD$ &\multicolumn{2}{c|}{\multirow{3}{*}{Fig.~\ref{fig:reduced43S}}} & \multicolumn{2}{c|}{\multirow{3}{*}{Fig.~\ref{fig:reduced43F}}}  & no & yes & yes &\\
 & $B|AD$ &   &   &   &  & no & yes & yes &\\
 & $D|AB$ &   &   &   &  & no$^{*}$ & yes / $\eta>0.07$ & yes &\\\cline{2-10}
 & $A|B|C$ & \multicolumn{2}{c|}{Fig.~\ref{fig:reduced43S}} & \multicolumn{2}{c|}{Fig.~\ref{fig:reduced43F}} & no & yes & N.A. &\\\cline{2-10}
 & $A|BC$ & \multicolumn{2}{c|}{\multirow{3}{*}{Fig.~\ref{fig:reduced43S}}} & \multicolumn{2}{c|}{\multirow{3}{*}{Fig.~\ref{fig:reduced43F}}}  & no & yes & yes &\\
 & $B|AC$ &   &   &   &  & no & yes & yes &\\
 & $C|AB$ &   &   &   &  & no$^{*}$ & yes / $\eta>0.07$ & yes &\\\cline{2-10}
reduced & $A|B$ & \multicolumn{2}{c|}{\multirow{6}{*}{Fig.~\ref{fig:reduced42f}}} & \multicolumn{2}{c|}{\multirow{6}{*}{Fig.~\ref{fig:reduced42f}}} &  no & yes & yes & \multirow{6}{*}{Fig.~\ref{fig:reduced2PPT}}\\
2 modes & $A|C$ &   &   &   &  & no & no & yes & \\
 & $A|D$ &   &   &   &  & no & no & yes &\\
 & $B|C$ &   &   &   &  & no & no & yes &\\
 & $B|D$ &   &   &   &  & no & no & yes &\\
 & $C|D$ &   &   &   &  & no & yes & yes &\\\hline
\end{tabular}
\caption{Summary of the full microscopic analysis of the entanglement structure. We show all partitions of the EPR ($N=2$), GHZ ($N=3$) and Cluster ($N=4$) states, as well as of all possible reduced density matrices, obtained by tracing over one or more modes. The table shows where the squeezing coefficient, the entanglement criterion based on the quantum Fisher information (QFI), and their associated optimal phase-space directions are presented. We further indicate whether a discontinuous behavior of the optimal phase-space directions is observed (column `disc.') for these partitions (where applicable we refer to a figure with further details). Notice
that some reduced partitions show a non-analytic transition from non-zero squeezing to zero squeezing as denoted by an asterisk ($^{*}$). Finally we display whether entanglement was detected by the criteria employed in this text and the PPT partition (only applicable for bi-partitions).}
\label{tab:summary}
\end{table*}

In addition to the data shown in the main manuscript, we provide a complete microscopic analysis of the entanglement structure of the three considered states. This encompasses an analysis of both entanglement coefficients, i.e., the squeezing coefficient and the Fisher information for all possible partitions and reduced density matrices, and an analysis of the optimal phase-space directions.

A summary of all partitions of the three states and their analysis is given in Tab.~\ref{tab:summary}. We analyzed a total of 44 partitions. Out of these, 12 are genuine multi-partitions, which can not be analyzed with the PPT criterion. In all of these cases, entanglement was revealed by the Fisher information. There are four bi-partitions, in which the squeezing coefficient and the Fisher information are unable to reveal entanglement, but its presence is demonstrated by the PPT criterion. Three partitions exhibit a discontinuous change of the optimal squeezing direction as a function of $\eta$. In three reduced partitions of the four-mode state, we detect entanglement with the Fisher information criterion for all $\eta$, while the squeezing coefficient is unable to detect it if $\eta$ is very small, otherwise both criteria coincide qualitatively.

The details of this analysis is presented in the following.

\subsection{Three-mode GHZ state}
Both entanglement coefficients for all four partitions of the three-mode GHZ state were shown in Fig. 3 in the main manuscript.

\subsubsection{Optimal phase-space directions and discontinuities of the coefficients}
Table~\ref{tab:GHZ} summarizes the phase-space directions $\mathbf{g}$ which maximize the entanglement coefficients. They identify a quadrature $\hat{q}(\mathbf{g})=\mathbf{g}\cdot\hat{\mathbf{r}}$ whose squeezing leads to maximal violation of the respective mode separability criterion (see main text). By construction, the coefficients do not depend on the overall sign of $\mathbf{g}$. 

We remark that the directions $\mathbf{g}$ which maximize violation of the separability criteria do not necessarily coincide with the quadrature that is most strongly squeezed, i.e., they are not always given by the minimal eigenvector of $\boldsymbol{\Gamma}_{\rho}$ or $\boldsymbol{\Omega}^T\boldsymbol{\Gamma}_{\Pi(\rho)}\boldsymbol{\Omega}$. Deviations between the two directions are found, e.g., for $\eta\neq 1$. Neither does the maximal eigenvalue $\lambda_{\max}$ necessarily identify a direction in phase space that maximizes the quantum Fisher information [instead of the difference with the local variances which is considered in Eq.~(4)]. If the goal is to maximize the Fisher information with a normalized phase space direction, the effect of noise can always be avoided by choosing a strongly squeezed (e.g. single-mode) quadrature that is not prone to losses.

\begin{table}[tb]
\begin{tabular}
[c]{|c|c|c|c|}\hline
partition &  $\eta\in$ & $\mathbf{g}_{\min}$ for Eq.~(7) & $\mathbf{e}_{\max}$ for l.h.s of Eq.~(4)\\\hline
& $[0,0.17]$ & $(0,0,1,0,-1,0)/\sqrt{2}$ & \multirow{2}{*}{$(0,0,1,0,-1,0)/\sqrt{2}$}\\
    \cline{2-3}%
$A|B|C$ & $[0.17,0.34]$ & \multirow{2}{*}{$(0,c_1,0,c_2,0,c_2)$} & \\\cline{2-2}\cline{4-4}
& $[0.34,1]$ & & $(0,c'_1,0,c'_2,0,c'_2)$\\\hline
$A|BC$ & $[0,1]$ & $(0,c_1,0,c_2,0,c_2)$ & $(0,c'_1,0,c'_2,0,c'_2)$\\\hline
$B|AC$ & $[0,1]$ & $(c_1,0,c_2,0,c_3,0)$ & $(c'_1,0,c'_2,0,c'_3,0)$\\\hline
$C|AB$ & $[0,1]$ & $(c_1,0,c_2,0,c_3,0)$ & $(c'_1,0,c'_2,0,c'_3,0)$\\\hline
\end{tabular}
\caption{Optimal phase-space directions to witness entanglement in the three-mode GHZ state. The coefficients $c_i$ usually vary with $\eta$ and are normalized to ensure that the phase-space direction is a unit vector.}
\label{tab:GHZ}
\end{table}

\begin{figure}[tb]
\includegraphics[width=.48\textwidth]{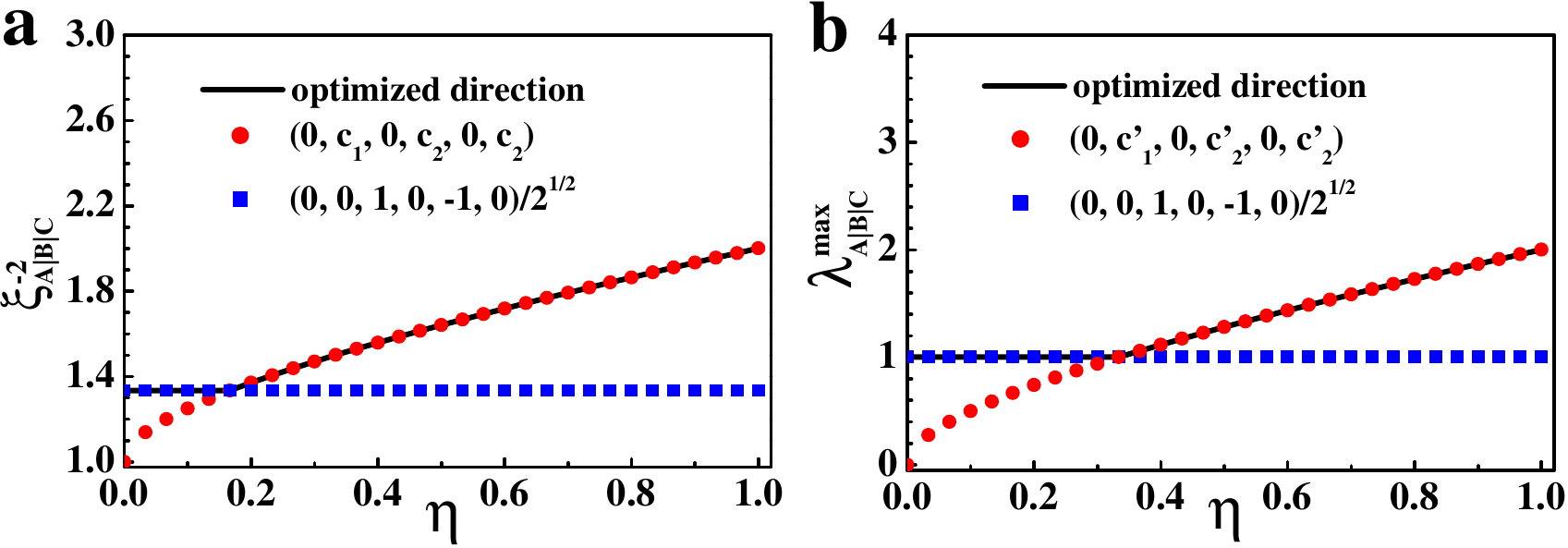}
\caption{Multi-mode squeezing coefficient (\textbf{a}) and Gaussian Fisher information entanglement criterion (\textbf{b}) for the multi-mode quadratures $(\hat{x}_B-\hat{x}_C)/\sqrt{2}$ (blue dots) and $c_1\hat{p}_A+c_2\hat{p}_B+c_2\hat{p}_C$ with optimized coefficients $c_1$ and $c_2$ (red dots), and maximized over all quadratures in the three-mode phase space (black lines). Abrupt changes of the maximal squeezing direction lead to discontinuous behavior of the entanglement coefficients as a function of $\eta$.}
\label{fig:AsBsC}
\end{figure}

Notably, we observe that the optimal direction for the $A|B|C$ direction changes abruptly at a critical value of $\eta$ due to the depletion of mode $A$. For strong losses, the strongest violation of separability is found for a phase-space direction with no overlap with mode $A$. This explains the discontinuous behavior of the two entanglement coefficients, as shown in Fig.~\ref{fig:AsBsC}.

\subsubsection{Reduced density matrices}
\begin{figure}[tb]
\includegraphics[width=.48\textwidth]{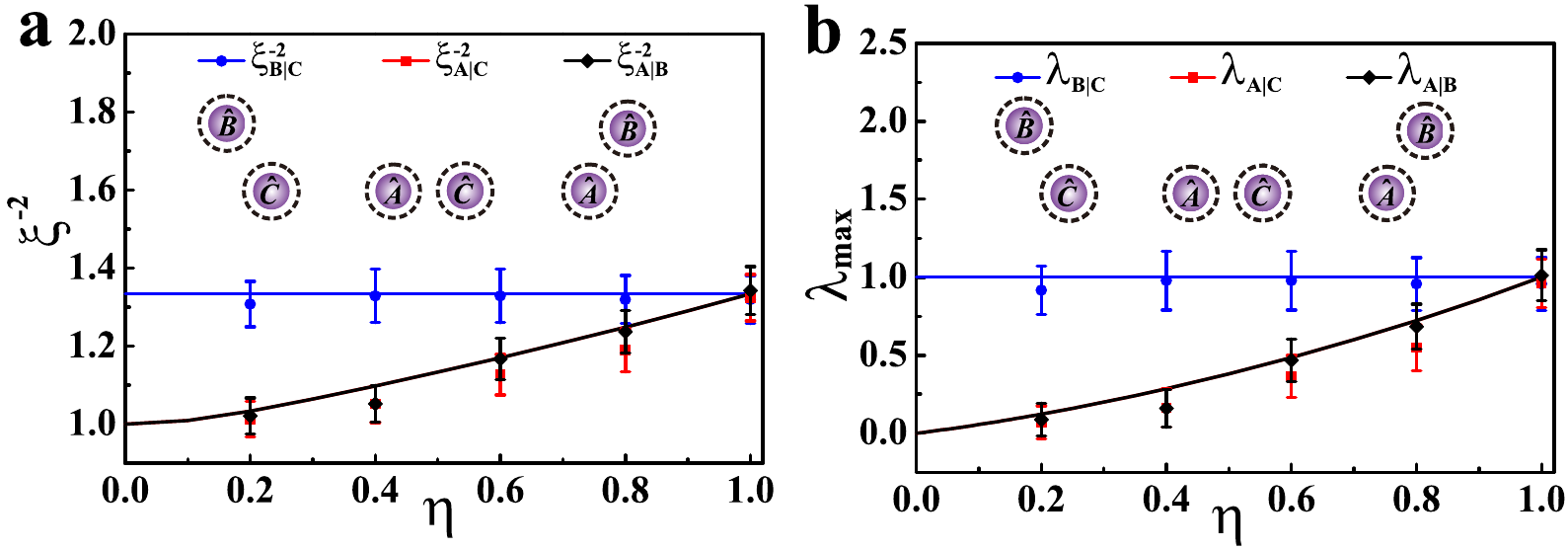}
\caption{Multi-mode squeezing coefficient (\textbf{a}) and Gaussian Fisher information entanglement criterion (\textbf{b}) for the reduced two-mode states of the three-mode GHZ state, obtained by tracing over one of the modes.}
\label{fig:reduced3}
\end{figure}

Further microscopic understanding of the entanglement structure can be gained by analyzing the reduced distributions after tracing out some of the modes. For the three-mode GHZ state, there are three different two-mode states, obtained by ignoring one of the three modes. These states can be analyzed using our entanglement coefficients in the same way as the two-mode state in the main manuscript. The coefficients are plotted in Fig.~\ref{fig:reduced3}.

When mode $A$ is ignored, the remaining state is completely invariant under losses of that mode (blue lines). The other two-mode reduced states for modes $A|B$ and $A|C$ show the same entanglement properties as the two-mode EPR state considered in the manuscript. This shows that the entanglement structure of the three-mode continuous-variable GHZ state differs vastly from that of GHZ states with discrete variables, which lose all of their entanglement as soon as one of the modes is trace out \cite{Dur}.

\subsection{Four-mode square cluster state}
\begin{figure}[tb]
\includegraphics[width=.48\textwidth]{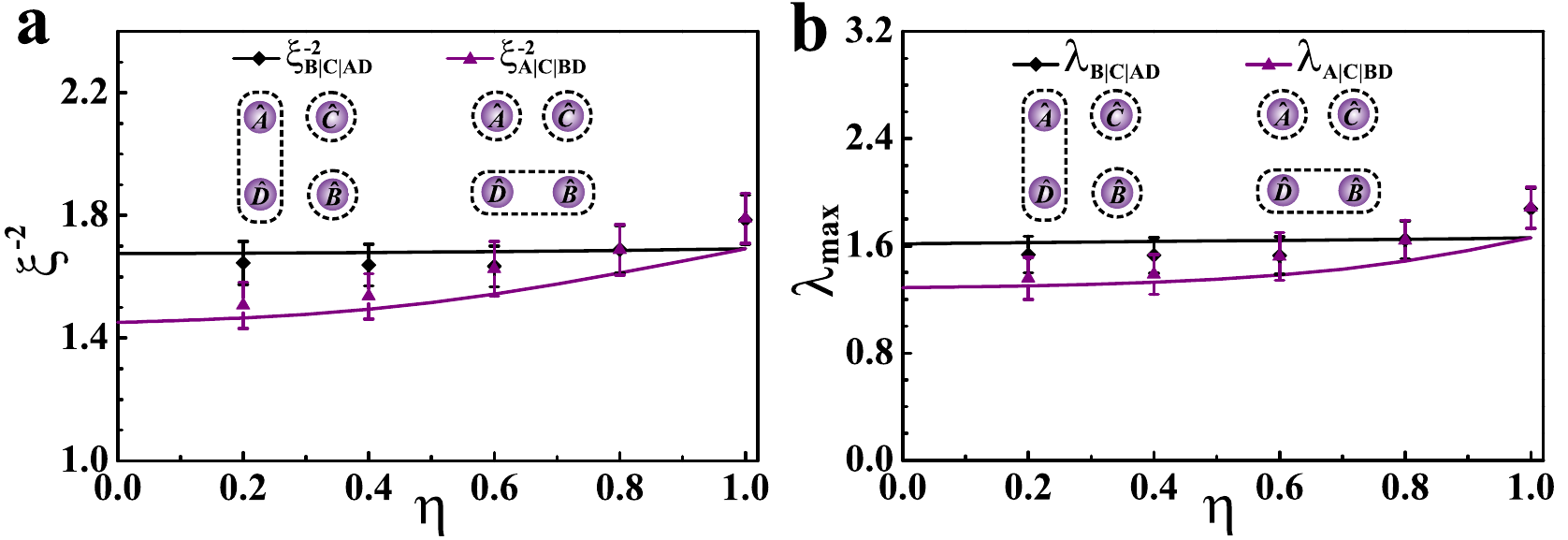}
\caption{Inverse multi-mode squeezing coefficients (\textbf{a}) and Gaussian Fisher information entanglement criterion (\textbf{b}) for $1\otimes1\otimes2$ partition of a CV four-mode Gaussian cluster state in a lossy channel, supplementing Fig.~4b in the main text.}
\label{fig:Cluster_Supp}
\end{figure}

In Fig.~\ref{fig:Cluster_Supp} we complete the analysis of the entanglement coefficients of Fig.~4 to all partitions by showing also the remaining data for the $B|C|AD$ and $A|C|BD$ partitions. It can be seen that the coefficients for $B|C|AD$ coincide with those of $B|D|AC$ shown in Fig.~4b and $A|C|BD$ coincides with $A|D|BC$. This is a consequence of the symmetric roles of modes $C$ and $D$.

\subsubsection{Optimal phase-space directions}
The optimal phase-space directions of the two entanglement criteria are shown for the $1\otimes 1\otimes 1\otimes 1$ partition in Tab.~\ref{tab:1111}, for the $1 \otimes 1 \otimes 2$ partitions in Tab.~\ref{tab:112}, for the $1 \otimes 3$ partitions in Tab.~\ref{tab:13}, and for the $2 \otimes 2$ partitions in Tab.~\ref{tab:22}. For simplicity, we only consider changes of the phase-space direction if the corresponding gain for the entanglement coefficient exceeds 1\% of the coefficient value.

\begin{table}[tb]
\begin{tabular}
[c]{|l|c|c|c|}\hline
partition & $\eta\in$ & $\mathbf{g}_{\min}$ for Eq.~(7) & $\mathbf{e}_{\max}$ for l.h.s of Eq.~(4)\\\hline
$A|B|C|D$ & $[0,1]$ & $(0,c_1,0,c_2,c_3,0,c_3,0)$ & $(0,c_1',0,c_2',c_3',0,c_3',0)$\\\hline
\end{tabular}
\caption{Optimal phase-space directions for the $1\otimes 1\otimes 1\otimes 1$ partition. The normalized coefficients $c_i$, $c_i'$ are optimized for each $\eta$.}
\label{tab:1111}
\end{table}

\begin{table}[tbp]
\begin{tabular}
[c]{|l|c|c|c|}\hline
partition &  $\eta\in$ & $\mathbf{g}_{\min}$ for Eq.~(7) & $\mathbf{e}_{\max}$ for l.h.s of Eq.~(4)\\\hline
& $[0,0.16]$ & $(0,c_1,0,c_2,c_3,0,c_3,0)$ & \multirow{2}{*}{$(0,c_1',0,c_2',c_3',0,c_3',0)$}\\\cline{2-3}%
$A|B|CD$ & $[0.16,0.31]$ & \multirow{2}{*}{$(c_1,0,c_2,0,0,c_3,0,c_3)$} & \\\cline{2-2}\cline{4-4}
& $[0.31,1]$ &  & $(c_1',0,c_2',0,0,c_3',0,c_3')$\\\hline
$A|D|BC$ & $[0,1]$ & $(c_1,0,c_2,0,0,c_3,0,c_4)$ & $(c_1',0,c_2',0,0,c_3',0,c_4')$\\\hline
$B|D|AC$ & $[0,1]$ & $(0,c_1,0,c_2,c_3,0,c_4,0)$ & $(0,c_1',0,c_2',c_3',0,c_4',0)$\\\hline
$C|D|AB$ & $[0,1]$ & $(0,c_1,0,c_2,c_3,0,c_3,0)$ & $(0,c_1',0,c_2',c_3',0,c_3',0)$\\\hline
\end{tabular}
\caption{Optimal phase-space directions for the $1\otimes 1\otimes 2$ partitions. The optimal directions of $B|C|AD$ and $B|D|AC$, as well as those of $A|C|BD$ and $A|D|BC$ coincide.}
\label{tab:112}
\end{table}

\begin{table}[tb]
\begin{tabular}
[c]{|l|c|c|c|}\hline
partition &  $\eta\in$ & $\mathbf{g}_{\min}$ for Eq.~(7) & $\mathbf{e}_{\max}$ for l.h.s of Eq.~(4)\\\hline
$A|BCD$ & $[0,1]$ & $(c_1,0,c_2,0,0,c_3,0,c_3)$ & $(c_1',0,c_2',0,0,c_3',0,c_3')$\\\hline
$B|ACD$ & $[0,1]$ & $(0,c_1,0,c_2,c_3,0,c_3,0)$ & $(0,c_1',0,c_2',c_3',0,c_3',0)$\\\hline
$C|ABD$ & $[0,1]$ & $(c_1,0,c_2,0,0,c_3,0,c_4)$ & $(c_1',0,c_2',0,0,c_3',0,c_4')$\\\hline
$D|ABC$ & $[0,1]$ & $(c_1,0,c_2,0,0,c_3,0,c_4)$ & $(c_1',0,c_2',0,0,c_3',0,c_4')$\\\hline
\end{tabular}
\caption{Optimal phase-space directions for the $1\otimes 3$ partitions.}
\label{tab:13}
\end{table}

\begin{table}[tb]
\begin{tabular}
[c]{|l|c|c|c|}\hline
partition &  $\eta\in$ & $\mathbf{g}_{\min}$ for Eq.~(7) & $\mathbf{e}_{\max}$ for l.h.s of Eq.~(4)\\\hline
$AB|CD$ & $[0,1]$ & $(0,c_1,0,c_2,c_3,0,c_3,0)$ & $(0,c_1',0,c_2',c_3',0,c_3',0)$\\\hline
$AC|BD$ & $[0,1]$ & $(0,0,c_1,c_2,c_3,c_4,c_5,c_6)$ & $(0,c_1',0,c_2',c_3',0,c_4',0)$\\\hline
$AD|BC$ & $[0,1]$ & $(0,0,c_1,c_2,c_3,c_4,c_5,c_6)$ & $(0,c_1',0,c_2',c_3',0,c_4',0)$\\\hline
\end{tabular}
\caption{Optimal phase-space directions for the $2\otimes 2$ partitions.}
\label{tab:22}
\end{table}

\begin{figure}[tb]
\includegraphics[width=.48\textwidth]{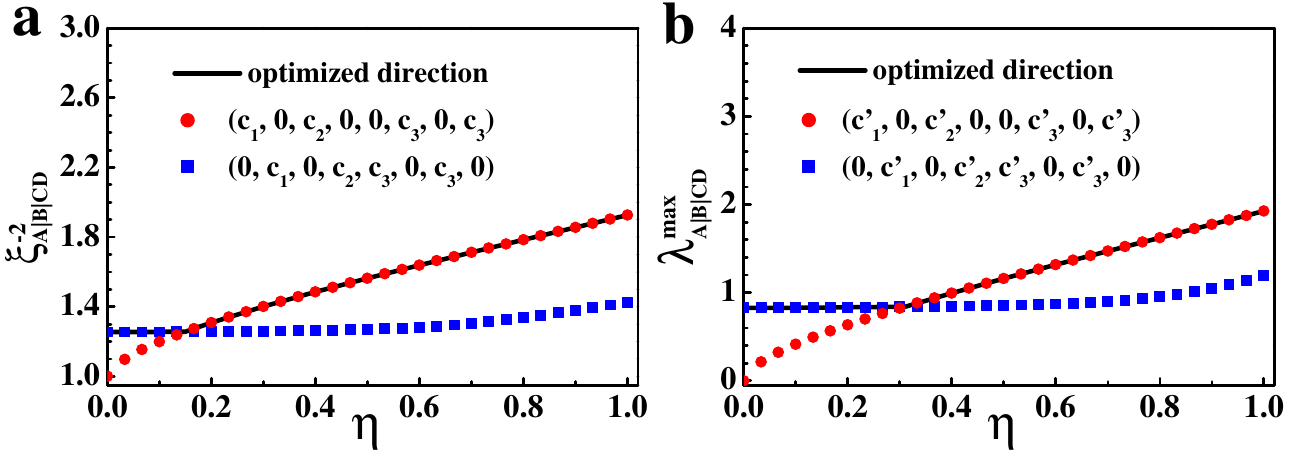}
\caption{Multi-mode squeezing coefficient (\textbf{a}) and Gaussian Fisher information entanglement criterion (\textbf{b}) for the partition $A|B|CD$ with multi-mode quadratures $c_1\hat{p}_A+c_2\hat{p}_B+c_3\hat{x}_C+c_3\hat{x}_D$ (blue dots) and $c_1\hat{x}_A+c_2\hat{x}_B+c_3\hat{p}_C+c_3\hat{p}_D$ with optimized and normalized coefficients $c_1,c_2,c_3$ (red dots), and maximized over all quadratures in the four-mode phase space (black lines).}
\label{fig:AsBsCD}
\end{figure}

The discontinuous behavior in the partition $A|B|CD$ [Fig.~4b and Fig.~4f] can again be explained by a transition of the optimal phase-space direction, see also Tab.~\ref{tab:112}. This is depicted in further detail in Fig.~\ref{fig:AsBsCD}.

\subsubsection{Reduced density matrices}

\begin{figure}[bt]
\includegraphics[width=.48\textwidth]{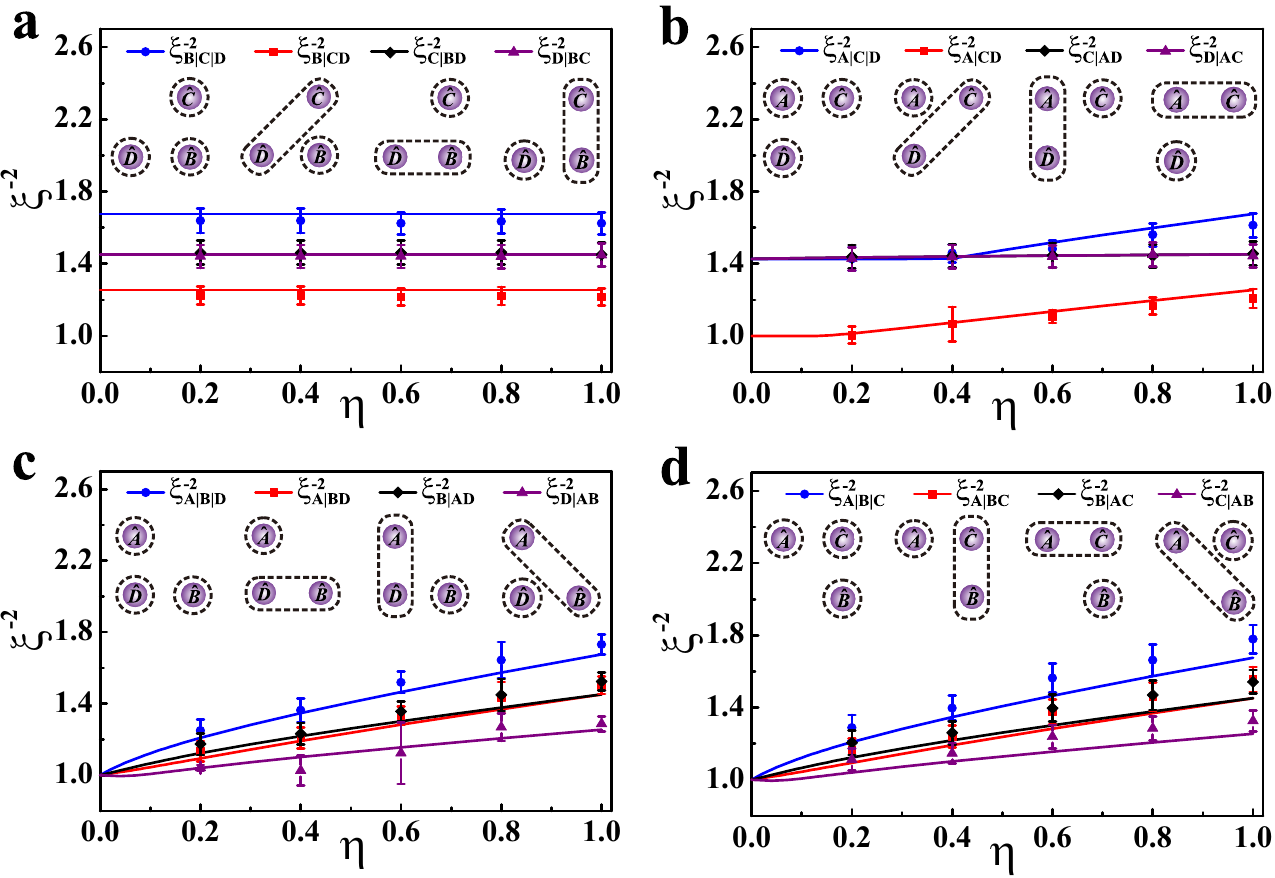}
\caption{Multi-mode squeezing coefficient for the reduced three-mode states of the four-mode square cluster state, obtained by tracing over one of the modes.}
\label{fig:reduced43S}
\end{figure}

\begin{figure}[bt]
\includegraphics[width=.48\textwidth]{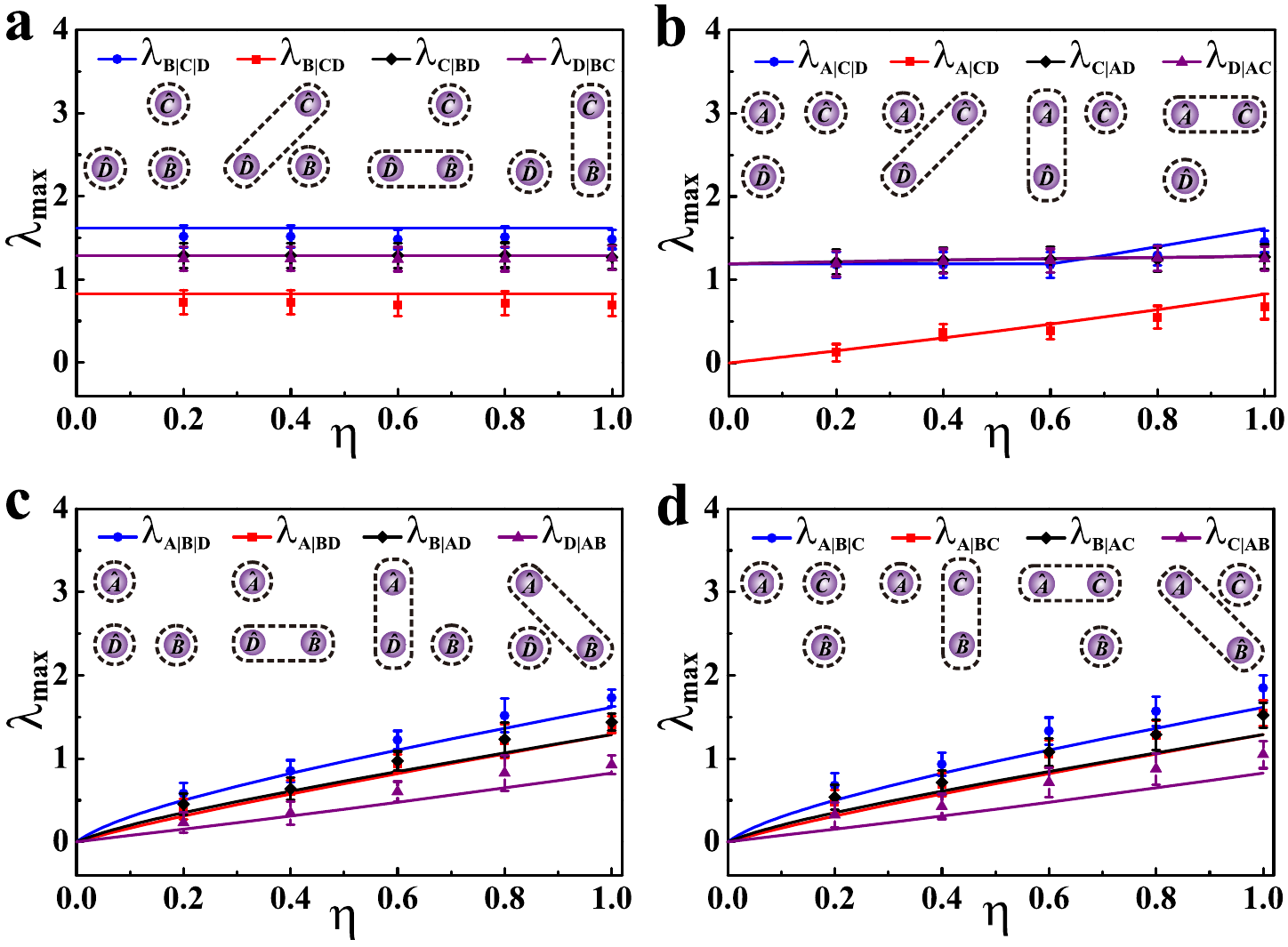}
\caption{Fisher information entanglement criterion for the reduced three-mode states of the four-mode square cluster state, obtained by tracing over one of the modes.}
\label{fig:reduced43F}
\end{figure}

\begin{figure}[tb]
\includegraphics[width=.48\textwidth]{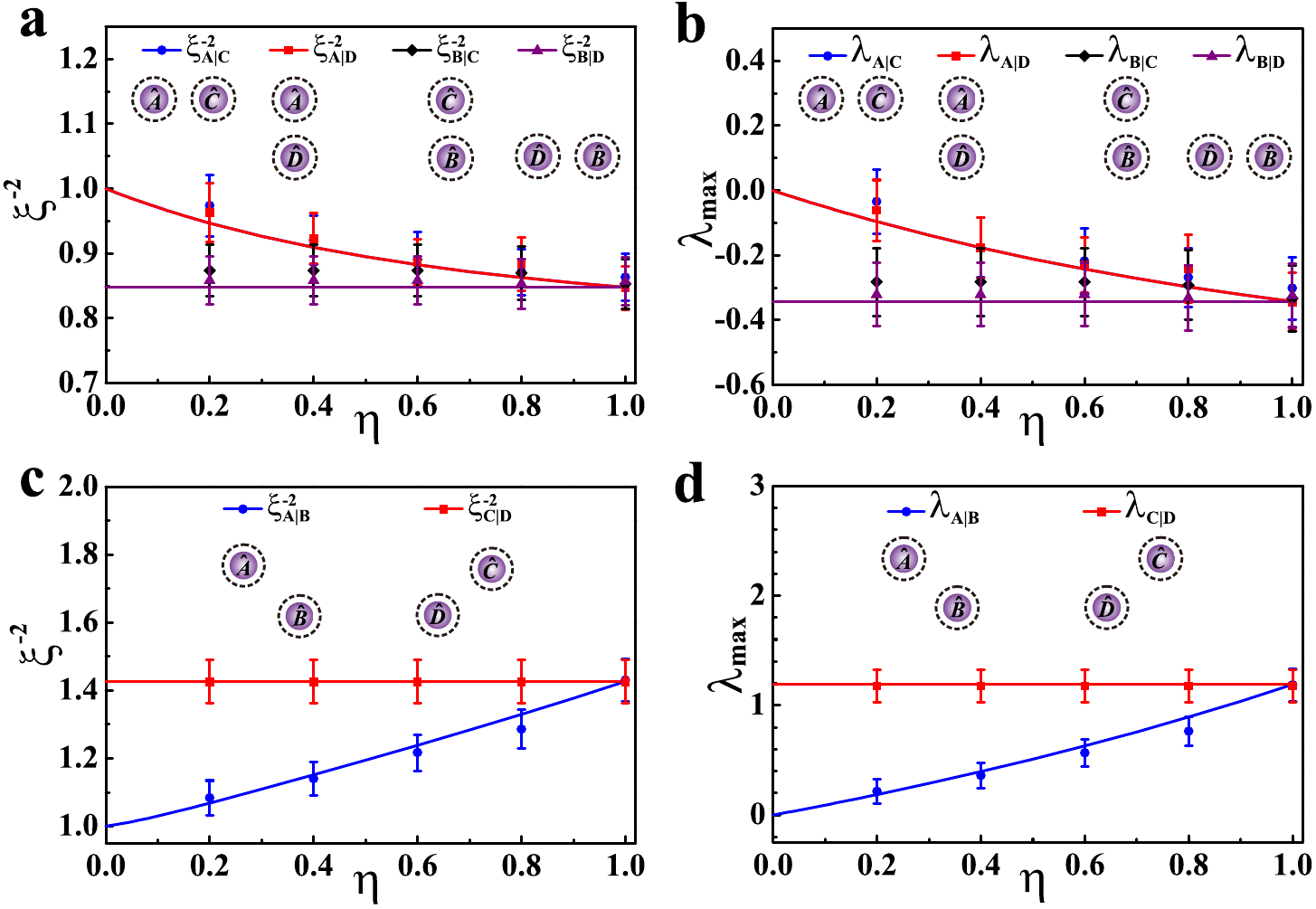}
\caption{Multi-mode squeezing coefficient and Gaussian Fisher information entanglement criterion for the reduced two-mode states of the four-mode square cluster state, obtained by tracing over two of the modes. No entanglement is witnessed in panels (\textbf{a}) and (\textbf{b}).}
\label{fig:reduced42f}
\end{figure}

The four-mode cluster state gives rise to a large ensemble of reduced density matrices with a rich entanglement structure. We show the squeezing coefficient [Fig.~\ref{fig:reduced43S}] and the Fisher information criterion $\lambda_{\max}$ [Fig.~\ref{fig:reduced43F}] for all partitions of the four reduced states of three modes, in analogy to the analysis of the three-mode CV GHZ state. We observe (i) the loss independence of modes $B$, $C$, and $D$ in subfigures (a), (ii) the symmetric roles of modes $C$ and $D$, as well as enhanced decoherence in absence of these modes in subfigures (c) and (d), (iii) an abrupt change of the optimal squeezing direction for the $A|C|D$ partition, and (iv) an entanglement structure akin to that of the three mode CV GHZ state after tracing out mode $B$ in subfigures (b).

Finally, we observe a difference between the squeezing coefficient and the Fisher information criterion. All states and all partitions contain entanglement for $\eta>0$, as is revealed by the Fisher information criterion, Fig.~\ref{fig:reduced43F}. In contrast, for small values of $\eta$, the squeezing coefficient is no longer able to detect the entanglement in the partitions $A|CD$, $C|AB$ and $D|AB$. The squeezing coefficient for $A|CD$ bends abruptly at $\eta=0.13$ and for smaller values no longer exceeds the separability threshold of $1$. This is analogous to the change of the optimal squeezing direction discussed in other cases before, with the difference that the optimal direction $\mathbf{g}_{\min}=(0, c1, c2, 0, c2, 0)$ for $\eta>0.13$ is unable to reveal entanglement when $\eta\leq 0.13$. In this case, a larger, yet still separable value of $\xi^{-2}_{A|CD}$ is achieved by ignoring mode $A$, i.e., $\mathbf{g}_{\min}=(0,0,0, 1,0 ,-1)/\sqrt{2}$. A similar phenomenon occurs for the partitions $C|AB$ and $D|AB$ around $\eta=0.07$, where several squeezing directions are almost degenerate below this value.

\begin{figure}[tb]
\includegraphics[width=.48\textwidth]{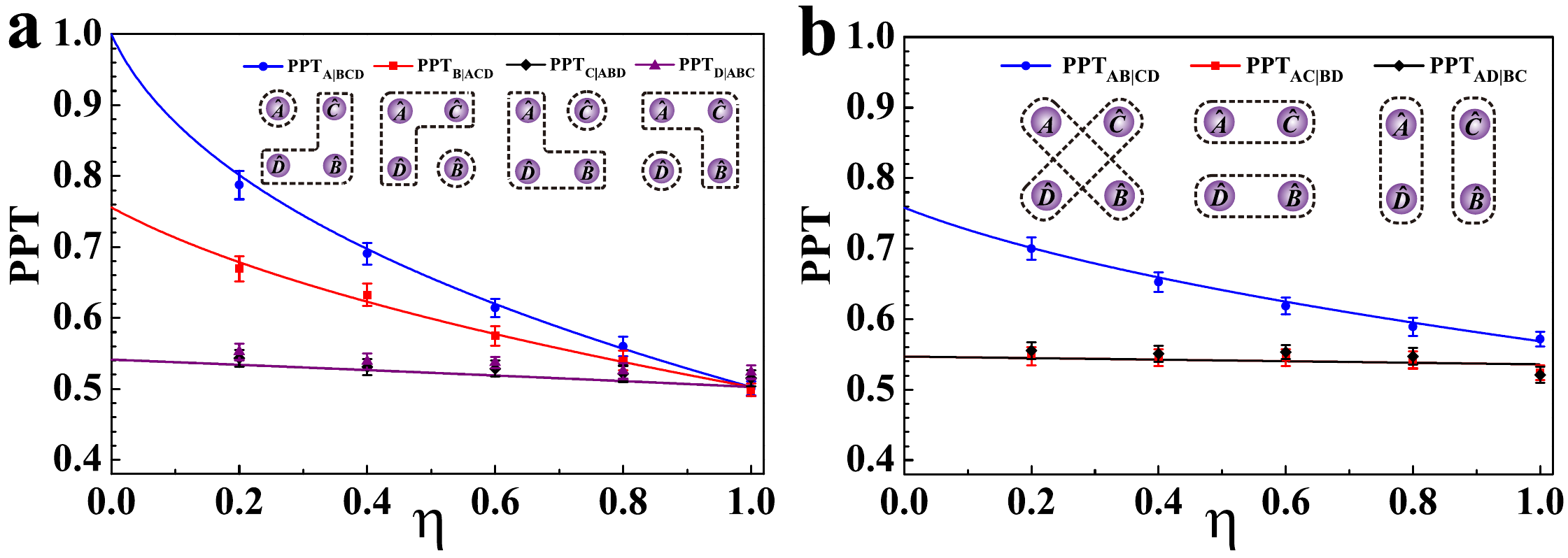}
\caption{PPT criterion for the four-mode CV cluster state. We plot the smallest symplectic eigenvalue of the covariance matrix of the partially transposed quantum state. This value is larger than one if and only if Eq.~(8) is satisfied. We show  $1\otimes 3$ partitions in (\textbf{a}) (in this case PPT is a necessary and sufficient condition for entanglement of Gaussian states) and $2\otimes 2$ partitions in (\textbf{b}). We identify entanglement for all values of $\eta\geq 0$ in all partitions, except at $\eta=0$ for $A|BCD$ when mode $A$ is isolated. This, as well as the qualitative dependence on $\eta$ is in complete agreement with the information provided by the two entanglement witnesses studied in our article, cf. Fig.~4 \textbf{c}, \textbf{d}, \textbf{g} and \textbf{h}.}
\label{fig:PPT}
\end{figure}
%\clearpage

The discrepancy between the two coefficients is due to the approximation that is made in the derivation of the squeezing coefficient. It is based on a Gaussian lower bound~(5) on the quantum Fisher information and, to ease the optimization procedure, the additional restriction to pairs of quadratures with maximal commutator, i.e., $\mathbf{h}=\boldsymbol{\Omega}\mathbf{g}$. Since the states considered here are still Gaussian, the limitation is due to the constrained optimization of quadratures.

We further show the entanglement witnesses for the reduced $1\otimes 1$ states, obtained from the four-mode cluster state after tracing over two modes in Fig.~\ref{fig:reduced42f}. Interestingly, our metrological entanglement criteria only reveal entanglement among the subsystems $A|B$ and $C|D$, i.e., the modes that are diagonal in the graph representation shown in Fig.~1c. However, an analysis with the PPT criterion reveals small amounts of entanglement also in the other bi-partitions (see Fig.~\ref{fig:reduced2PPT} below). Hence, these highly mixed reduced states represent examples of Gaussian states whose entanglement is not revealed by comparing the metrological sensitivity with their separability bounds. The comparison with the PPT criterion will be discussed further in the next section.

\begin{figure}[tb]
\includegraphics[width=.49\textwidth]{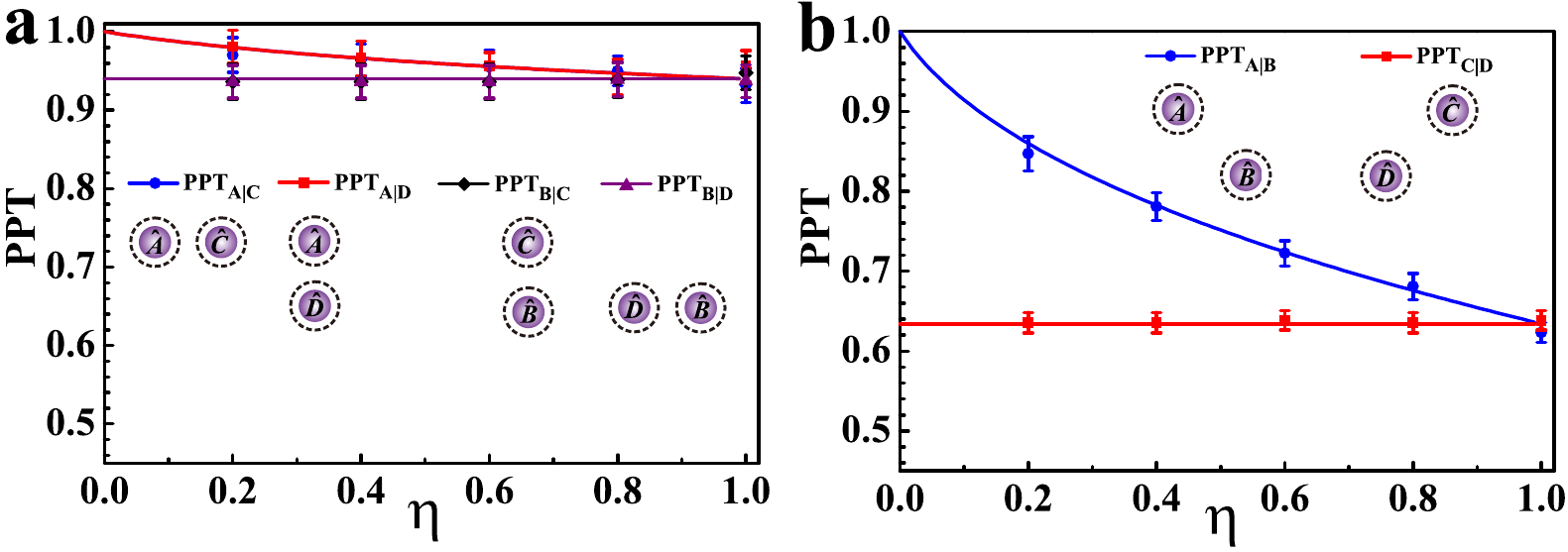}
\caption{PPT criterion for the reduced two-mode states of the four-mode cluster state. The strongest entanglement is found for the diagonal modes (\textbf{b}), but small amounts of entanglement can be found also for neighboring modes (\textbf{a}).}
\label{fig:reduced2PPT}
\end{figure}

\begin{widetext}
\begin{figure*}[htbp]
\includegraphics[width=.98\textwidth]{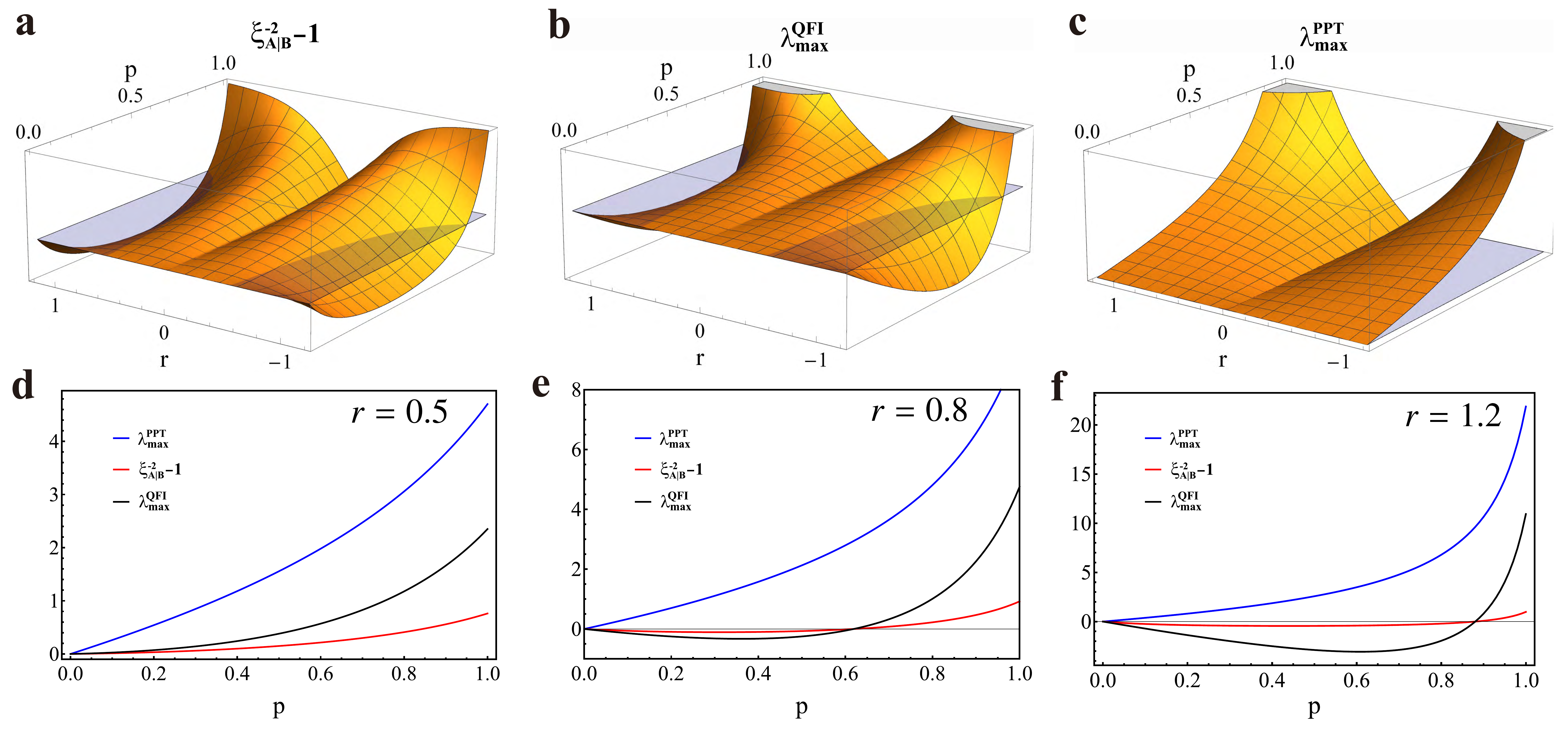}
\caption{Inverse multi-mode squeezing coefficient~(7), Gaussian quantum Fisher information criterion obtained from the largest eigenvalue of the l.h.s of Eq.~(4), and the PPT criterion, formulated as the largest eigenvalue of the l.h.s. of Eq.~(8) for the two-mode Gaussian entangled states mixed with vacuum with covariance matrix~(\ref{eq:GammaPPT}) as a function of the squeezing $r$ and the mixing weight $p$. All quantities are plotted in a normalized way such that values above zero indicate entanglement. In the 3D plots in the upper row, the blue semi-transparent plane indicates the zero value. The PPT criterion (\textbf{c}) is necessary and sufficient since the state is Gaussian and consists of two modes. The other two criteria derived from the Fisher information (\textbf{a}) and (\textbf{b}) are unable to detect the entanglement when $|r|>\mathrm{arccoth}(2)\approx 0.549$, see text for details. Direct comparisons for the values of $r=0.5$ (\textbf{d}), $r=0.8$ (\textbf{e}), and $r=1.2$ (\textbf{f}) are displayed in the lower row.}
\label{fig:PPTSQZ}
\end{figure*}
\end{widetext}

\section{Benchmarking the entanglement witnesses with the PPT criterion}

In the special case where a single subsystem is separated in a bi-partition from the rest of the modes, the PPT criterion becomes a necessary and sufficient condition for separability of Gaussian states \cite{WernerWolf}. It therefore represents an ideal benchmark for the novel entanglement witness considered in our work. 

In summary, our criteria show qualitative agreement with the PPT criterion for all bi-partitions for all three states of $N=2$, $N=3$ and $N=4$. However, differences are observed if all reduced distributions are taken into account. The metrological tools do not reveal entanglement in some of the $1\otimes 1$ partitions of the reduced two-mode states, obtained by tracing over two of the modes of the cluster state, while the PPT criterion still indicates entanglement. In these states, entanglement is present, but the achievable metrological sensitivity stays below their associated separability limit. 

\subsection{Analysis of the experimental data with PPT}
As an example, we show the PPT criterion for the $1\otimes 3$ partitions of the four-mode CV square cluster state in Fig.~\ref{fig:PPT}a. We find indeed that our entanglement coefficients [Fig.~4c and Fig.~4g] show the same qualitative behavior as the PPT condition. 

In fact, whenever our criteria detect entanglement in a $1\otimes (N-1)$ partition, the PPT criterion necessarily also detects it. Such a direct conclusion cannot be drawn for the $2\otimes 2$ partitions of the four-mode cluster state, as the PPT is no longer a necessary and sufficient criterion for entanglement of Gaussian states. However, also in this case the PPT criterion detects the entanglement (see Fig.~\ref{fig:PPT}b).

We analyzed the data from all generated quantum states with the PPT criterion for comparison and find the same qualitative behavior in all bi-partitions and subsystems except for some of the reduced $1\otimes 1$ states represented in Fig.~\ref{fig:reduced42f}. 

These examples show that our criteria are not necessary and sufficient for separability of Gaussian states and therefore not equivalent to the PPT criterion in general. 

\begin{figure}[tb]
\includegraphics[width=.49\textwidth]{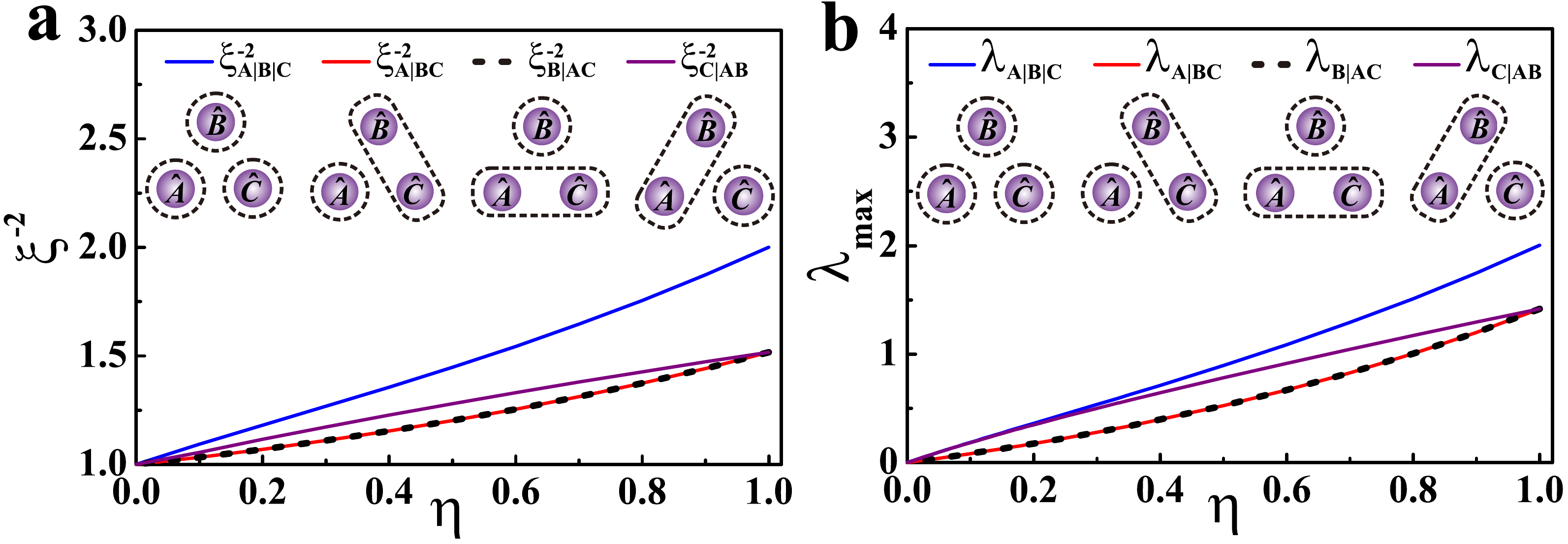}
\caption{Theoretical results for the three-mode GHZ state when modes $A$ and $B$ transmit through lossy channels with transmission efficiency $\eta$. \textbf{a} Inverse multi-mode squeezing coefficients. \textbf{b} Gaussian Fisher information entanglement criterion.}
\label{fig:GHZABloss}
\end{figure}

\subsection{A class of PPT-entangled states not detected by squeezing and Fisher information}

We further illustrate this inequivalence by another family of entangled Gaussian states. Consider the CV two-mode Gaussian entangled states (without losses) incoherently mixed with the vacuum state. These states are Gaussian and have a covariance matrix
\begin{align}\label{eq:GammaPPT}
\boldsymbol{\Gamma}=\frac{p}{2}\begin{pmatrix} c & 0 & s & 0\\
0& c & 0 & -s \\
s & 0 & c & 0 \\
0&-s&0&c
\end{pmatrix}+\frac{1-p}{2}\begin{pmatrix} 1 & 0 & 0 & 0\\
0& 1 & 0 & 0 \\
0 & 0 & 1 & 0 \\
0&0&0&1
\end{pmatrix},
\end{align}
with $c=\cosh(2r)$ and $s=\sinh(2r)$ and $r$ quantifies the squeezing. The coefficient $p$ determines the relative weight. Incoherent decay into the vacuum transforms an initial two-mode Gaussian entangled state into the above state with $p=e^{-\gamma t}$, where $\gamma$ is the decay rate and $t$ is the evolved time \cite{Paris}.

\begin{figure*}[tb]
\includegraphics[width=.98\textwidth]{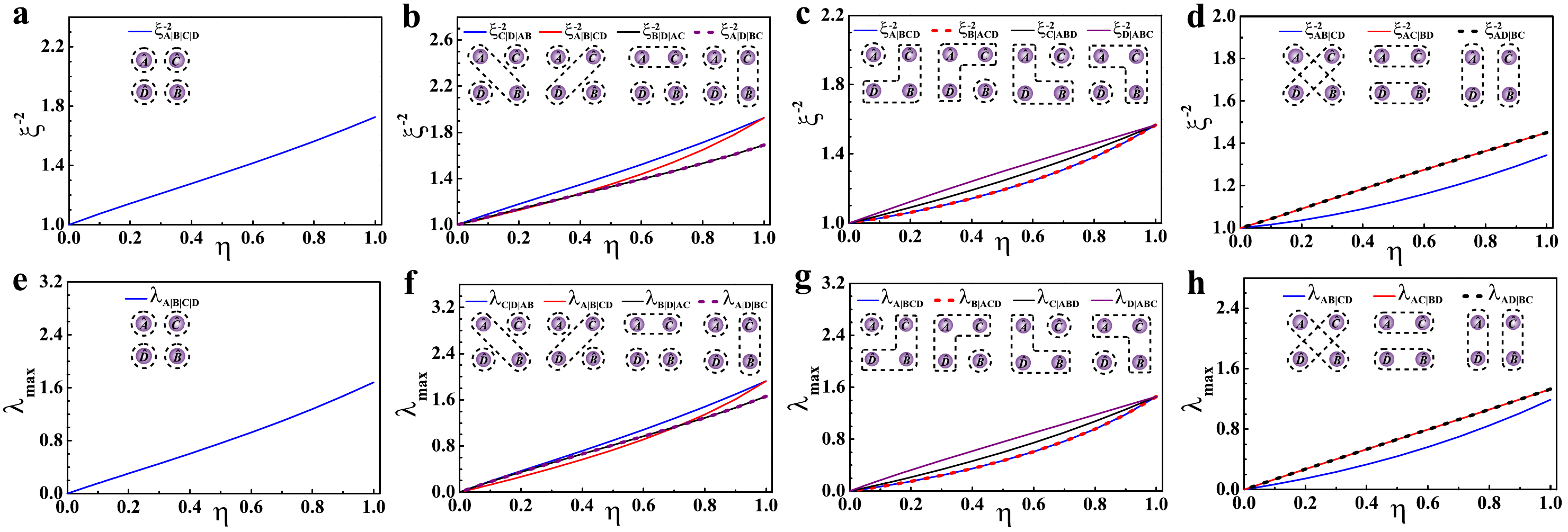}
\caption{Theoretical results for the four-mode Gaussian cluster state when modes $A$, $B$, and $C$ transmit through lossy channels with transmission efficiency $\eta$. \textbf{a--d} Inverse multi-mode squeezing coefficients $\xi^{-2}$ for the partitions of classes $1\otimes1\otimes1\otimes1$, $1\otimes1\otimes2$, $1\otimes3$, and $2\otimes2$, respectively. \textbf{e--h} The corresponding Gaussian Fisher information entanglement criterion.}
\label{fig:ClusterABCloss}
\end{figure*}

The coefficients studied in this paper only reveal the entanglement of this state for $|r|\leq r_0=\mathrm{arccoth}(2)\approx 0.549$ over the entire range of $p$, while it is entangled for any value of $r\neq 0$ and $p\neq 0$, as shown by the PPT criterion, see Fig.~\ref{fig:PPTSQZ}. When $|r|>r_0$, entanglement remains unrevealed by the squeezing criterion~(7) and the Fisher information~(4) in the range $p\in[0,p_{\max}]$ with 
\begin{align}
p_{\max}=\frac{1}{2}(2-\coth|r|)(1+\coth|r|),
\end{align}
as displayed in Fig.~\ref{fig:PPTSQZ}. This result can be analytically obtained by applying the criterion~(4) to the state characterized by the covariance matrix~(\ref{eq:GammaPPT}).

\section{Effect of losses on multiple modes}
In realistic quantum communication schemes, it is more common that more than one mode of a multipartitie CV entangled state suffer from losses. Here, we theoretically investigate the entanglement for CV three-mode GHZ state and four-mode Gaussian cluster state when losses are added on two modes and three modes, respectively. For simplicity, we only consider the case where the amount of loss added on each mode is the same.

The squeezing coefficient, as well as the Gaussian Fisher
information entanglement criterion, are plotted in Fig.~\ref{fig:GHZABloss} for all four partitions of a three-mode GHZ state when modes $A$ and $B$ both have a non-unit transmission efficiency $\eta$. Fig.~\ref{fig:ClusterABCloss} shows the result for four-mode cluster state when modes $A$, $B$ an $C$ have a non-unit transmission efficiency $\eta$. Although entanglement for all the partitions decrease faster as $\eta$ decreases compared with the case when loss is only added on mode $A$, entanglement for all different partitions always exists unless $\eta$ decreases to 0, which further confirms the robustness of CV entangled states.

\end{document}